\documentclass[%
 reprint,
 amsmath,amssymb,
 aps,
]{revtex4-2}

\usepackage{graphicx}
\usepackage{dcolumn}
\usepackage{bm}

\usepackage{epsfig} 
\usepackage{float}
\usepackage{xcolor}
\usepackage{epstopdf}
\usepackage[labelfont=bf,textfont=md]{caption}
\usepackage[hidelinks]{hyperref}

\newcommand{\cc}[1]{\textcolor{black}{#1}}
\newcommand{\dd}[1]{\textcolor{black}{#1}}

\newenvironment{ddcomment}{\par\color{green}}{\par}

\begin{document}

\preprint{APS/123-QED}

\title{Buckling Metamaterials for Extreme Vibration Damping}

\author{David M.J. Dykstra}
 \email{dmj.dykstra@gmail.com}
\author{Coen Lenting}%
\author{Alexandre Masurier}%
\author{Corentin Coulais}%
 \email{coulais@uva.nl}
\affiliation{%
Institute of Physics, University of Amsterdam,  Science Park 904, 1098 XH, Amsterdam, the Netherlands\\
}%


\begin{abstract}
Damping mechanical resonances is a formidable challenge in an increasing number of applications.
Many of the passive damping methods rely on using low stiffness dissipative elements
, complex mechanical structures 
or electrical systems
, while active vibration damping systems typically add an additional layer of complexity. 
However, in many cases, the reduced stiffness or additional complexity and mass render these vibration damping methods unfeasible. 
%
This article introduces a new method for passive vibration damping by allowing buckling of the primary load path, \cc{which sets an upper limit for vibration transmission: the transmitted acceleration saturates at a maximum value, no matter what the input acceleration is.
This nonlinear mechanism 
%
leads to an extreme damping coefficient $\tan \delta \approx 0.23$ in 
a metal metamaterial---orders of magnitude larger than the linear damping of traditional lightweight structural materials.
%
This article demonstrates this principle} experimentally and numerically in \cc{free-standing rubber and metal} mechanical metamaterials
\cc{
over a range of accelerations, and shows that bi-directional buckling can further improve its performance.}
\cc{Buckling metamaterials pave the way towards extreme vibration damping without mass or stiffness penalty, and as such could be applicable in a multitude of high-tech applications, including aerospace structures, vehicles and sensitive instruments.}
\end{abstract}

\maketitle

\

\begin{figure*}[t!]
\includegraphics[width=1.\textwidth]{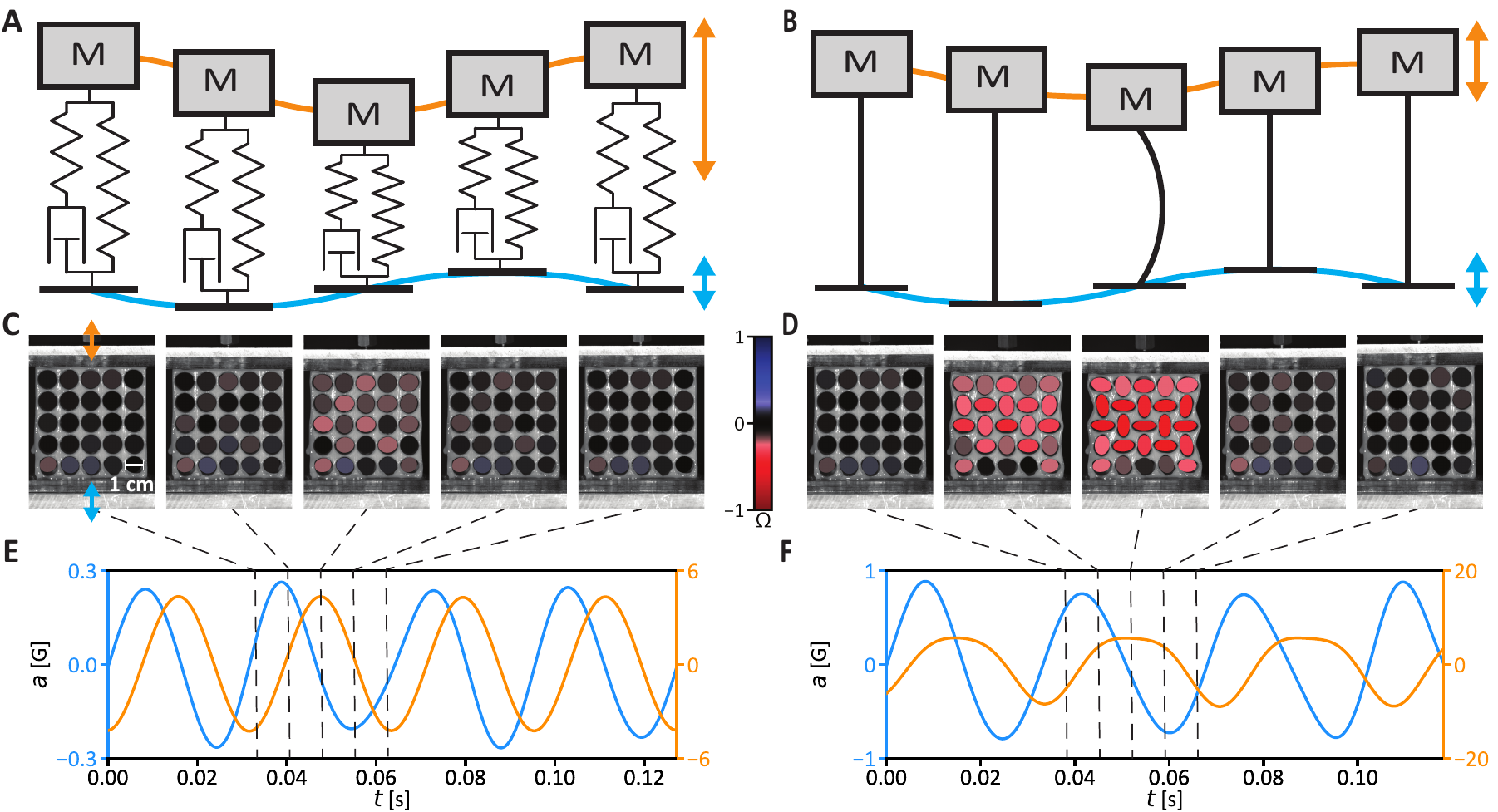}
\caption{\textbf{Damping vibrations with buckling.} (A) A mass ($M$) spring damper system, with base excitation (blue) can show a large amplified response (orange) around resonance. 
(B) When the spring is a slender beam, which can buckle when subjected to a sufficient compressive load from the base excitation, the amplified response may be lower. (C,D) show the deformation of a holar sample with mass mounted on top when subjected to a base excitation from the bottom around the eigenfrequency. (C) corresponds to a base excitation acceleration of 0.26 G at 33.8 Hz, while (D) corresponds to a base excitation acceleration of 0.89 G at 33.0 Hz. The ellipticity of the holes, $\Omega$, is tracked with red and blue ellipses (Section~\ref{sec:image}, color bar). (E,F) Base excitations (blue) of 0.26 G (E) and 0.89 G (F) induce output accelerations (orange) of 4.3 G (E) and 5.7 G (F) respectively.}
\label{fig1}
\end{figure*}

Any mechanical system will exhibit a resonance. At this resonance, the transmission of force and acceleration is maximal. Limiting the amplification of acceleration is paramount in a wide range of applications where vibrations can cause unwanted noise and failure.
A paradigmatic example is that of a mass and spring damper shaken from the bottom (Fig. \ref{fig1}A): at resonance, the mass will vibrate with a much higher acceleration than the input acceleration provided by the shaker. While myriad strategies using highly viscoelastic materials~\cite{zhou2016research,baz2019active}, negative stiffness components~\cite{lakes2001extreme,wang2004extreme,jaglinski2007composite,kochmann2017exploiting,li2020negative,balaji2021applications}, band-gap metamaterials~\cite{lu2009phononic,hussein2010band,hussein2013metadamping,wang2014harnessing,cummer2016controlling,krushynska2017coupling,gao2022acoustic} and active control~\cite{baz2019active,collette2011review,aridogan2015review} have been proposed, they typically suffer from added mass or loss in stiffness.

Here, we propose to use Euler buckling as a functional mechanism to create vibrations absorbers (Fig. \ref{fig1}B): buckling structures are simultaneously stiff thanks to their high stiffness prior to buckling, yet limit the transmission of acceleration under post-buckling since they exhibit a force plateau under compression. Although the idea is best illustrated with a single column under axial vibrations (Fig. \ref{fig1}B), such a structure would collapse in post-buckling and would not be suitable for applications that require freestanding unsupported structures. 
This is precisely where metamaterials could come to the rescue, as they can combine buckling in one direction and structural stiffness in the other directions~\cite{florijn2014programmable,restrepo2015phase,Shan_AdvMat2015,frenzel2016tailored,yuan20193d,dykstra2019viscoelastic,hector2023energy}. While their shock response has been well studied, their nonlinear vibration response remains poorly understood.

In this article, we demonstrate that 
free-standing load-carrying metamaterials that undergo a buckling instability 
provide an upper limit for transmission of vibrations.
%
We show that this efficient vibration absorption stems from elastic and damping nonlinearities that are induced by buckling. We further generalize the concept to metallic metamaterials and to metamaterials that buckle both under compression and tension. We obtain extreme damping with respect to ordinary lightweight structural materials. Our work  demonstrates that buckling metamaterials are a competitive solution for lightweight structures combining high damping and high specific stiffness in high-tech applications.

To create a structure that can maintain its own lateral stability in post-buckling, we first turn to one of the most common designs in flexible mechanical metamaterials~\cite{grima_auxetic,Mullin_PRL2007,Bertoldi_NatRevMat}: a polymeric slab that is patterned with a square array of circular holes (Fig. \ref{fig1}CD). This metamaterial exhibits a global buckling mode, where the pattern of pores becomes an array of ellipses with alternating orientations (Fig. \ref{fig1}D, centre). To determine its nonlinear vibration characteristics, we mount a mass on top and subject the sample to a base excitation around the eigenfrequency at 2 different levels: a low level of 0.26 G (Fig. \ref{fig1}CE) and a high level of 0.89 G (Fig. \ref{fig1}df), where G=9.81~m/s$^{2}$ is the acceleration of gravity. 

At the lower excitation level, we observe that the holes remain close to circular (Fig. \ref{fig1}C\dd{, see also Supplementary Video 1}) and we observe both a sinusoidal output response, which has a $\pi / 2$ phase lag with respect to the input excitation (Fig. \ref{fig1}E). This was to be expected based on linear vibrations~\cite{inman1994engineering}. At the higher excitation level, the sample buckles as seen in Fig. \ref{fig1}D\dd{, see also Supplementary Video 1}. As a result, the output response in Fig. \ref{fig1}F is no longer perfectly sinusoidal. Importantly, despite the input level increasing by more than a factor three (from 0.26 G to 0.89 G), the output level in compressive direction merely changes by a third (from 4.3 G to 5.7 G). More surprisingly, the maximum acceleration in tensile direction also only increases by a factor of two instead of three (from 4.3 G to 8.9 G). This suggests that compressive buckling also dampens vibrations in tensile direction.

\begin{figure*}[t!]
\centering
\includegraphics[width=1\textwidth]{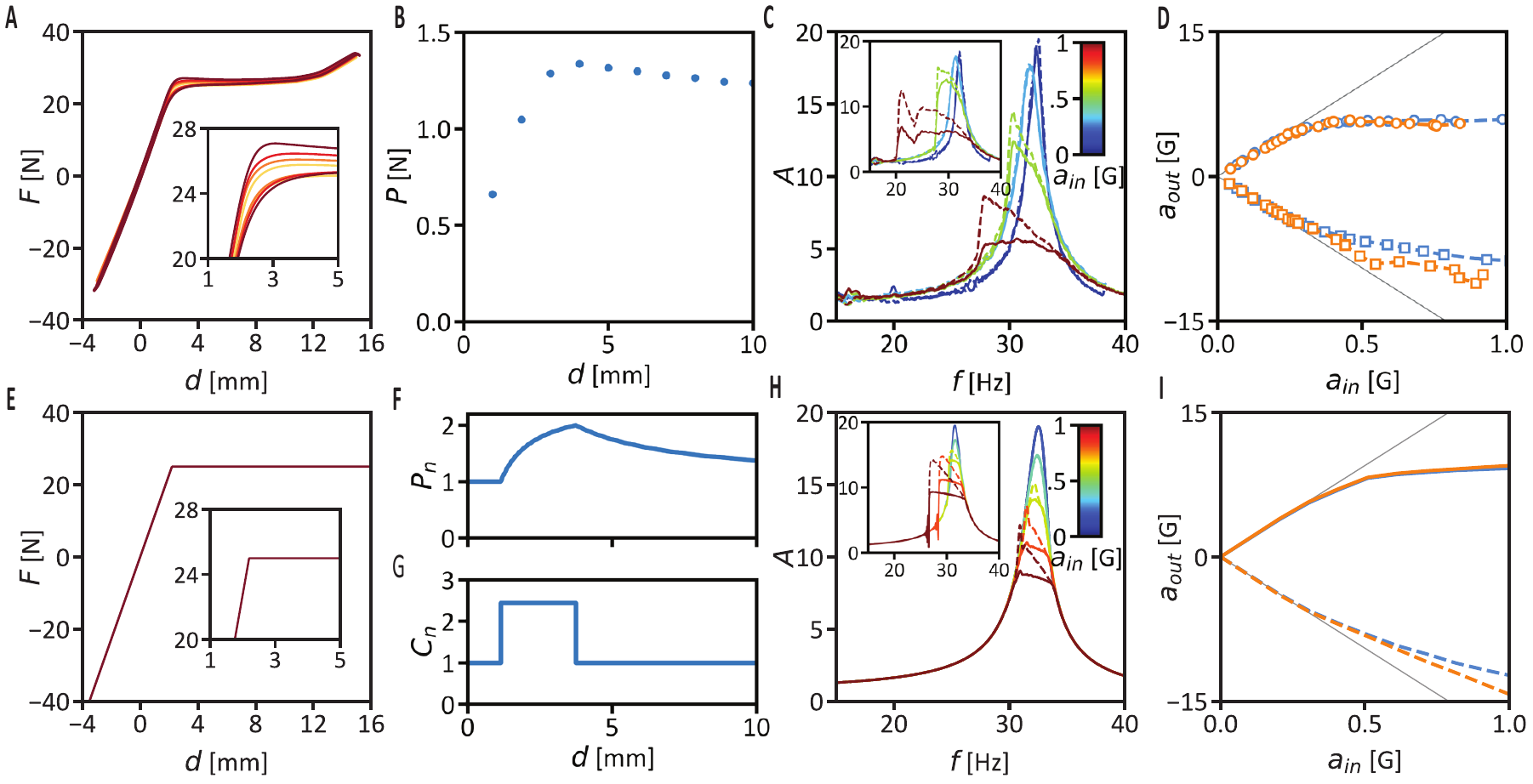}
\caption{\textbf{Vibration damping performance of buckling metamaterials.} (A) Force ($F$)  displacement ($d$) curves during compression-tension tests of the sample of Fig. \ref{fig1}DE. The four curves in D and E correspond to loading rates of 1, 10, 100 and 1000 mm/min from yellow to red respectively. (E) Equivalent force-displacement curve of simplified model (see Section~\ref{sec:app_num}). (B,F) Average hysteretic force over compression range when performing tension-compression experiments at 1000 mm/min: (B) experimental, (F) numerical, normalized with the average hysteretic force for 0.1mm compression-tension. (G) Numerical Voigt damper strength as function of the compressive displacement, normalized by the linear Voigt damper strength (Methods). (C,H) Maximum acceleration amplification factor $A$ as function of frequency $f$ during a frequency sweep with rising and (inset) dropping frequency: (C) experimental and (H) numerical. Solid lines correspond to tension while dashed lines correspond to compression. The color of the curve indicates the input acceleration $a_{in}$. (D,I) Maximum output acceleration $a_{out}$ across the frequency range: (E) experimental and (I) numerical. Blue (orange) curves correspond to rising (dropping) frequencies. Circles and solid lines (squares and dashed lines) correspond to compression (tension) in E and I respectively. Grey lines show the linearized trend. }
\label{fig2}
\end{figure*}

This efficient vibration damping stems from nonlinearities induced by buckling. To quantify such nonlinearities, we perform compression and tension mechanical tests at various strain rates (Fig. \ref{fig2}A, see Methods for details). As expected~\cite{Mullin_PRL2007,bertoldi2010negative}, while the response is nearly linear in tension and for compression less than $2$ mm, a buckling instability occurs at a compressive displacement of 2 mm. This instability induces a force plateau,
a key nonlinearity that explains the saturation of acceleration under compression seen in Fig.~\ref{fig1}F. However, this nonlinearity alone does not suffice to explain the reduction of acceleration in the tensile direction. 

The missing ingredient is an additional damping nonlinearity that is also rooted in buckling. Indeed, when we vary the loading rate, we observe that the amount of hysteresis increases significantly around the point of buckling when we increase the loading rate from 1 mm/min to 1000 mm/min (Fig. \ref{fig2}A-inset).
To better quantify this effect, we compress and extend the sample at a loading rate of 1000 mm/min up to different compression levels and measure the average hysteretic force
(difference between loading and unloading) across the loading regime as function of the compression. 
We see that the hysteresis is non-monotonic with a maximum at a compression of 4 mm, which corresponds to the buckling point. This non-monotonic damping differs from that of linear viscoelastic materials (e.g. 
the Voigt damper in Fig. \ref{fig1}A), where the average hysteretic force is constant. One concludes that buckling amplifies viscoelastic effects. This can be interpreted by the fact that the material that makes up the slender parts of the metamaterial undergoes much larger strain rates than the full structure does.  Moreover, another key component is present specifically in vibrations with a base excitation. Effectively speaking, nonlinearities break resonance. As the peak force levels do not increase linearly with acceleration exciting level, the energy injected in the system does not increase linearly  with acceleration exciting level either. 


These combined elastic and damping nonlinearities both team up to efficiently dissipate vibrations. To quantify such dissipation, we measure the amplification factor (ratio between output and input acceleration) as a function of frequency: we perform frequency sweeps at various input levels with both rising (Fig. \ref{fig2}C) and dropping (Fig. \ref{fig2}C-inset) frequency levels, passing the resonance. 
For rising frequencies, buckling is very efficient at limiting the vibration transmission. For small excitations (blue in Fig. \ref{fig2}C), amplification factors up to 19 are found. During post-buckling the peak amplification factors across the frequency domain become as low as 6 in compression (bordeaux solid lines) and 9 in tension (bordeaux dashed lines). \dd{This corresponds to more than doubling and tripling the amount of damping for tension and compression respectively.} The peak amplification factors also shift to lower frequencies. For dropping frequencies, we experience a different trend. We observe larger amplification factors (Fig. \ref{fig2}C-inset) at smaller frequencies compared to a frequency sweep with a rising frequency. This difference between rising and dropping frequency sweeps demonstrates bistability post-buckling at frequency ranges below that of the linear resonance. 
Equivalently, this reduction of the amplification factor corresponds to a saturation of the maximum 
output acceleration, $a_{out}$, across the frequency range as function of the input acceleration, $a_{in}$ (Fig. \ref{fig2}D). In the compressive direction (circles), we observe a hard limit on the maximum vibration transmission at 5.9 G. In tension (squares), we observe that the trend becomes markedly lower than the linear trendline (grey) post-buckling. This behavior is consistent with analytical examples of vibrations of mass-spring dampers with softening springs such as quadratic or cubic examples \cite{kovacic2011duffing,brennan2008jump,guckenheimer2013nonlinear}. However, while analytical examples of vibrations of mass-spring dampers with weakly nonlinear softening springs typically demonstrate increased amplification factors at larger excitations~\cite{kovacic2011duffing,brennan2008jump}, Fig. \ref{fig2}C demonstrates the opposite: the drastic nonlinearities spawned by buckling shave off the resonance peak and efficiently limit vibration transmission. 

However, the fact that buckling dampens effectively simple frequency sweeps does not yet guarantee that it dampens more complex vibrations. After all, nonlinear vibration responses can not be linearly combined like linear vibration responses can be combined. Therefore, we also subject the sample of Fig. \ref{fig1} to random vibrations (See Section \ref{app_random}) and we find again that the maximum transmission of acceleration saturates at 6.0 G. This demonstrates that buckling based vibration damping works effectively regardless of the type of vibration: controlled or random.

In order to design buckling-based vibration damping, it is also necessary to be able to predict it. In theory, finite element methods could be used to model the response of buckling based vibration damping. However, nonlinear dynamic finite element methods are notoriously expensive computationally, especially for a very large number of cycles~\cite{dykstra2022extreme}. For this purpose, we develop a simple numerical model, based on a nonlinear mass spring damper, similar to the simplified representation of Fig. \ref{fig1}A. However, instead of a linear model, we use nonlinear force-displacement  (Fig. \ref{fig2}E) and dashpot strength-displacement (Fig. \ref{fig2}FG) curves. We tune these parameters to fit the experimental results of Fig. \ref{fig2}AB and the low excitation eigenfrequency of Fig. \ref{fig2}C (Methods). We then subject the model numerically to the same frequency sweeps as Fig. \ref{fig2}CD (Methods) and obtain the equivalent numerical results in Fig. \ref{fig2}HI. 

We observe that the results in Fig. \ref{fig2}HI match the results of Fig. \ref{fig2}CD well: not only qualitatively but also quantitatively. The only significant difference found is that the numerical model predicts slightly higher output accelerations post-buckling than the experiments. This could suggest that the bilinear approximations of Fig. \ref{fig2}EG oversimplify the nonlinear dissipation of the experiments. 
However, the fact that the numerical predictions are slightly higher than the experimental results also show that the numerical results slightly underestimate the damping performance. Moreover, while the fitting parameters of Fig. \ref{fig2}EG have been obtained based on the experiments (See Section~\ref{sec:app_num}), these could also be generated using nonlinear finite element methods, without modeling the entire frequency sweep tests with these same finite element methods. Together, this implies that buckling metamaterials with target stiffness and target damping could conservatively be designed and predicted using a very simple single element numerical model.

\section{High stiffness materials}

\begin{figure*}[ht!]
\centering 
\includegraphics[width=1.\textwidth]{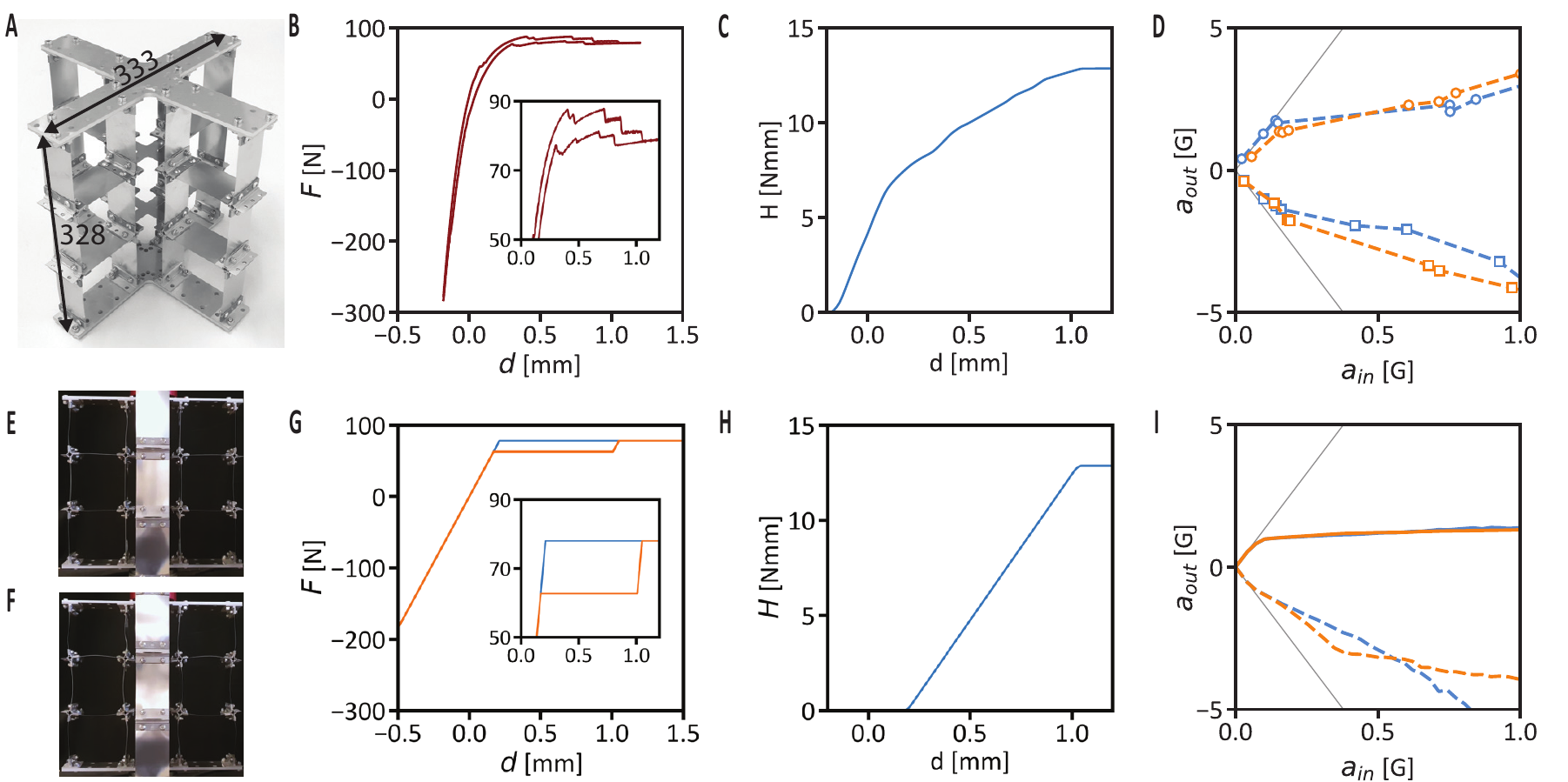}
\caption{\textbf{Metallic buckling metamaterial for vibration damping.} (A) Metallic metamaterial consisting of two crossing $3 \times 3$ unit cells, allowing buckling, \dd{with side view: (E) unbuckled and uncompressed, (F) buckled and compressed by 1.2mm}. 
(B) Force displacement curve of metal metamaterial with zoomed insert. (C) Cumulative dissipated energy, $H$, corresponding to (B). (D) When subjected to a base vibration, with a 1.4kg mounted on top: as function of the average input acceleration $a_{in}$, the peak output acceleration $a_{out}$ across the frequency domain. Orange (blue) curves correspond to rising (dropping) frequencies. Circles (squares) correspond to compression (tension). Grey lines show the average linearized trend. (G-I) Numerical results for simulations, equivalent to the experimental results of B-D (See Methods).}
\label{fig:metal}
\end{figure*}

So far, we have demonstrated that vibration damping using buckling metamaterials made out of an elastomer is efficient. However, our results 
suffer from three main shortcomings. First of all, elastomers inherently suffer from a low specific stiffness, which makes them inherently unsuited for load bearing structures. Second, it is far from obvious that buckling metamaterials can be generalized to 
stiff materials, such as metals, fibre reinforced composites or ceramics, which have a low yield or failure strain~\cite{Ashby2016}. This implies that stiff metamaterials cannot undergo repeated thick-walled buckling. Finally, while viscoelastic buckling will increase the amount of dissipation (Fig. \ref{fig1}D (inset)), the resulting extra amount of dissipated energy can still be relatively small compared to the total strain energy in the system for base materials with low damping coefficients.

To overcome the first two problems, we can use thin-walled metal metamaterials. We can prevent yielding while buckling by letting the entire mechanism bend, as is demonstrated in Fig. \ref{fig:metal}EF, as opposed to localized mechanism deformation in Fig. \ref{fig1}D. Furthermore, while straight thin-walled beams inherently exhibit a low buckling load, the buckling load can be increased and tailored using curvature about the longitudinal axis of each beam. 
In turn, by using the thin-walled design of Fig. \ref{fig:metal}EF, the design becomes inherently sensitive to shear. We can remove this compliance to shear by creating a 3D structure made out of two copies of the 2D mechanism, wherein the two 2D mechanism stabilize one another. 

We construct the metamaterial sample of Fig. \ref{fig:metal}AEF. This sample consists of 0.15 mm thin laser cut steel sheets (AISI 301 Full Hard) of 90$\times$50 mm, bolted to 3D printed aluminium connectors (AlSi10Mg, selective laser sintering). The steel sheets are pre-curved by 1.5 mm. The top and bottom crosses are made from 5 mm thick CNC milled aluminium, while the cross sheets in the middle are made from 0.25 mm thin laser cut steel sheets (AISI 301 Full Hard). The geometry has been selected with the help of static nonlinear finite element methods in Abaqus. 
The sample measures 333 $\times$ 333 $\times$ 328 mm and has a mass of 1.9 kg.

When we compress the sample slowly (1 mm/min) and plot the force-displacement curve in Fig. \ref{fig:metal}B, we observe a buckling force plateau at 78 N. While buckling, we observe snap-through instabilities (Fig. \ref{fig:metal}B inset). When the pre-curved members buckle, they snap to an uncurved state about their longitudinal axis. In particular, we observe that snap-through releases strain energy and excites local vibration modes of the metal sheets. This immediately helps to solve the third problem: insufficient dissipation for low damping base materials, which was identified in the beginning of this section. The snap-through instabilities show negative stiffness and dissipate energy~\cite{lakes2001extreme,lakes2001extreme_PRL,jaglinski2007composite}. 

In turn, this dissipated energy as function of the displacement is plotted in Fig. \ref{fig:metal}C. It is also noteworthy that we observe that the force-displacement curve before buckling is not entirely linear due to imperfections in the sample.

Similar to what we did for the elastomeric sample in Fig. \ref{fig1} and Fig. \ref{fig2}, we add a mass of 1.5 kg on the top of the sample and subject the metal sample to a vibrational base excitation\dd{, see also Supplementary Video 2}. We subject the sample to frequency sweeps from low to high frequencies and back at several acceleration excitation levels (See Section~\ref{subsec:app_defining}). We track the acceleration response at the base and top of the sample, similar to what we did for the elastomeric sample in Fig. \ref{fig2}. We also compute the maximum output acceleration across the frequency domain as function of the input acceleration level in Fig. \ref{fig:metal}D. Again, we find that buckling based vibration damping works. More specifically, the metal sample features a loss coefficient $\tan \delta \approx 1/ A_{max} \approx 0.23$ at $a_{in} = 1$ G, \dd{a tripling of the damping coefficient with respect to the low excitation response.} . This loss coefficient greatly surpasses those of traditional light-weight structural materials such as aluminium alloys ($1 \cdot 10^{-4} - 2 \cdot 10^{-3}$), steels ($2 \cdot 10^{-4} - 3 \cdot 10^{-3}$) or carbon fibre reinforced polymers ($1 \cdot 10^{-3} - 3 \cdot 10^{-3}$)~\cite{Ashby2016}. This shows that buckling based vibration damping can be used to surpass the Ashby limits of loss coefficient versus specific modulus~\cite{Ashby2016}. 

Moreover, it is noteworthy that the metal sample did not suffer from fatigue damage during the performed vibration tests. Over the course of the test campaign, the metal sample has been subjected to around $\approx 10^{5}$ vibration cycles, across a variety of frequency and acceleration levels. At no point during this testing campaign has any visual damage been identified. This is all the more impressive considering that snap-through induces local modes, which may even vibrate at a higher frequency. This shows that it is also possible to produce buckling metamaterials for vibration damping with a high specific stiffness without being fatigue sensitive. 

Similar to what we did for the elastomeric sample, we can also model the performance of the metal sample using a numerical model in Fig. \ref{fig:metal}GHI (See Section~\ref{sec:app_num}). We can approximate the force-displacement curve of Fig. \ref{fig:metal}B using the bilinear force-displacement curve of Fig. \ref{fig:metal}G (blue). Here, we omit the imperfections in the force-displacement curve of Fig. \ref{fig:metal}B to see how a sample with little imperfections would respond. However, unlike what we considered for the numerical model of the elastomeric sample in Fig. \ref{fig2}EFG, we will omit nonlinear viscosity and instead only consider the dissipation due to the snap-through instabilities. To do this, we smear the hysteresis of Fig. \ref{fig:metal}B linearly over the range between the start of buckling in Fig. \ref{fig:metal}G and the distance of the last snap-through instability in Fig. \ref{fig:metal}B. As such, we obtain the unbuckling force-displacement curve of Fig. \ref{fig:metal}G (orange), with corresponding hysteresis-displacement curve in Fig. \ref{fig:metal}H. We then subject the model numerically to the same back-and-forth frequency sweeps as the experimental experimental of Fig. \ref{fig:metal}A and obtain the equivalent numerical results of Fig. \ref{fig:metal}D in Fig. \ref{fig:metal}I. Again, we also find that buckling based vibration damping works numerically, particularly in compression. 

We do however obtain some significant differences between the experiments and numerics. Namely, the numerics predict lower peak accelerations in compression than the experiments and higher peak accelerations in tension, especially for dropping frequencies. The lower peak acceleration in compression could be because the metal sample also has additional local modes, which the numerical model does not take into account. Furthermore, viscoelasticity can delay buckling, which can increase the peak load~\cite{dykstra2019viscoelastic}. A similar effect in turn explains why the numerical mode predicts higher accelerations in tensile direction: delayed buckling and unbuckling induced by viscoelasticity can increase dissipation~\cite{dykstra2019viscoelastic}, which can lead to lower amplification factors, as discussed above in the case of the elastomeric sample.
Despite these discrepancies, even without considering nonlinear damping, but when considering snap-through induced dissipation, the numerical model shows valid trends. This demonstrates that such a simple numerical model can still be used to predict and design the performance of buckling based vibration damping metamaterials, even when they are made from thin-walled structures with high specific moduli and snap-through instabilities.

\section{Bi-directional buckling}

\begin{figure*}[ht!]
\centering 
\includegraphics[width=1.\textwidth]{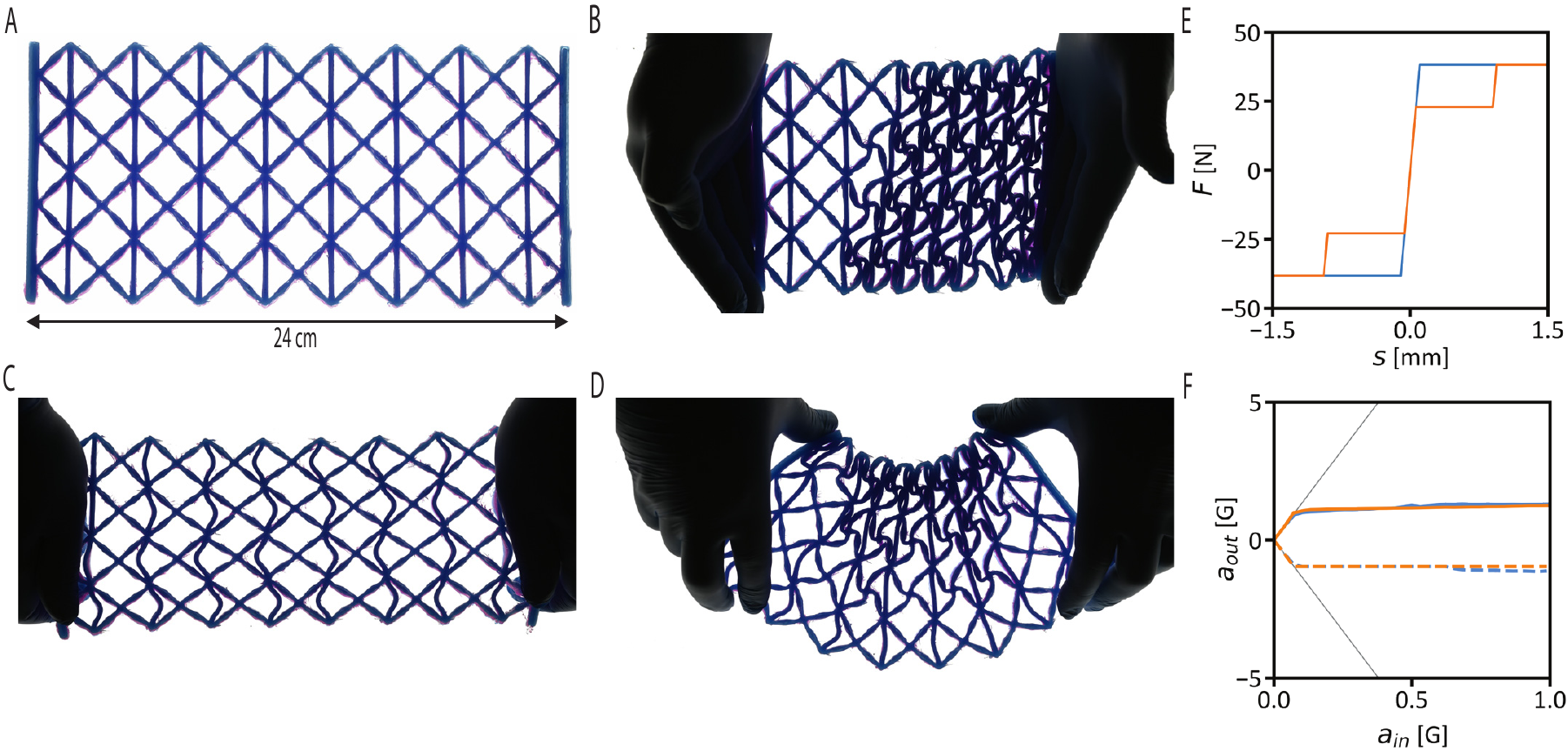}
\caption{\textbf{Buckling based vibration damping in tension and compression}. (A) 3D printed TPU sample allowing buckling in two directions, buckling respectively in (B) compression, (C) tension and (D) bending. (E) Force-displacement for featuring tensile and compressive snap-through buckling, based on the force-displacement curve of Fig. \ref{fig:metal}G. (F) For numerical up and down frequency sweeps at various acceleration levels, the maximum output acceleration $a_{out}$ across the frequency range. Blue (orange) curves correspond to rising (dropping) frequencies. Solid (dashed) lines correspond to compression (tension). Grey lines show the linearized trend.}
\label{fig:bidir}
\end{figure*}
																																																																																																																																																																																											
So far, we have demonstrated that buckling metamaterials dampen vibrations for both soft and stiff materials. However, all of the cases analyzed so far consider buckling exclusively in the compressive direction. Yet, it is also possible to realize geometries that can buckle in multiple directions, such as the geometries considered in Fig. \ref{fig:bidir}. In Fig. \ref{fig:bidir}A, we have created a sample, which can buckle in both the compressive (Fig. \ref{fig:bidir}B) and the tensile (Fig. \ref{fig:bidir}C) direction\dd{ (See section \ref{subsec:MultiFabri} for manufacturing details)}. 

This sample has in fact been optimized to buckle in tension and compression at the same strain level using Bayesian optimisation and nonlinear finite element methods with Python and Abaqus~\cite{masurier2021thesis}. 
The sample can even buckle simultaneously in tension and compression when subjected to more complex loading cases, such as bending in Fig. \ref{fig:bidir}D.

While this sample demonstrates that buckling can be achieved simultaneously in both tension and compression, it does not offer sufficient stiffness by itself to be used as a load bearing vibration damping structure. However, we can generalize our numerical model to simulate the vibration response of structures that offer snap-through buckling in both tension and compression. To do so, we adjust the model we used to simulate Fig. \ref{fig:metal}G-I, and apply the force-displacement curve of Fig. \ref{fig:bidir}E instead. This force-displacement curve is in fact the symmetrically buckling version of Fig. \ref{fig:metal}G, with the numerically considered mass subtracted (See Section~\ref{sec:app_num}). We then use our model to subject the sample numerically to the same up and down frequency sweeps at different base accelerations as we did for Fig. \ref{fig:metal}I. When we track the maximum output acceleration across the frequency domain, $a_{out}$ as function of the input acceleration, $a_{in}$, we obtain the results of Fig. \ref{fig:bidir}F. Here, we find that buckling based vibration damping is very effective to set an upper limit for vibration transmission. In compression, it is already more effective than what was the case when buckling only occurred in compressive direction, as in Fig. \ref{fig:metal}I. This is because the snap-through induced hysteresis per cycle is twice as large, as snap-through occurs in both compression and tension. Furthermore, Fig. \ref{fig:bidir}F demonstrates that the same upper level for vibration transmission can be set in tensile direction, making it much more effective in tensile direction than what was the case in Fig. \ref{fig:metal}I. This shows that buckling metamaterials offering buckling in both tension and compression can offer increased vibration damping over those that only buckle in compression.

\section{Outlook}
Of course, the geometry of Fig. \ref{fig:bidir}A-D is not the only metamaterial design, which can offer buckling in both tensile and compressive direction. For instance, kirigami forms a common alternative~\cite{rafsanjani2017buckling,vella2019buffering,chen2020kirigami}. However, if we look carefully at the structure of Fig. \ref{fig:bidir}A, we can see that it is very similar to lightweight lattice structures~\cite{smeets2021structural,hunt2022review}. Such fibre-reinforced lattice structures are specifically designed for their high strength and stiffness to weight ratio and can be made in both 2D or 3D geometries. Typically in such structures, the members have been sized such that they do not buckle under load~\cite{smeets2021structural,hunt2022review}. However, their members could also be sized instead to allow for elastic buckling, possibly accompanied by snap-through, such as the sample shown in Fig. \ref{fig:metal}. \dd{A specific advantage of such lattice structures, but also of the metal sample of Fig. \ref{fig:metal}, is that they are stretching dominated pre-buckling. This implies that a much higher specific stiffness can be achieved than in many bending dominated structures~\cite{deshpande2001foam,restrepo2015phase,Shan_AdvMat2015,frenzel2016tailored,hector2023energy}.}  As such, it may be possible to produce buckling based vibration damping metamaterials with a very high stiffness to weight ratio. 

\dd{Furthermore, while the current research demonstrates a threefold improvement in damping for both samples, we expect that much larger improvements in damping coefficient could be obtained in systems with large amplification factors at small excitations. More specifically, if the sample of Fig. \ref{fig:metal}, or another similar structure produced out of high stiffness materials, had less imperfections, we would expect a much larger amplification at small excitation. Meanwhile, we would not expect the amplification factors of excitations that induce post-buckling to increase significantly. As such, the increase in damping performance of buckling metamaterials for vibration damping could even be much larger than reported in this manuscript.} Finally, while this \dd{manuscript} has focused on limiting the transmission of a specific resonance, the method itself does not depend on a frequency. In fact, this method should work to restrict transmissions of any vibration. This could be a big advantage in cases with multiple modes.


Open challenges remain however. Specifically, it still has to be demonstrated that buckling based vibration damping metamaterials can be produced with an even higher specific stiffness, without significant imperfections. Similarly, the possible ceiling limits of performance still need to be identified. 

When these points are addressed, we anticipate applications in any situation, where a high specific stiffness is required along with high damping, including aerospace, sensitive instruments and high-tech machinery~\cite{baz2019active,balaji2021applications,diez2019review,yan2017shunt,gripp2018vibration,zhou2016research}.

\clearpage
\section{Materials and Methods}

\subsection{Sample design and fabrication}

\subsubsection{Elastomeric holar sample fabrication}

To create the sample of Fig. \ref{fig1}CD, we pour a two component silicone rubber (Zhermack Elite Double 32) in a 3D printed mold, manufactured with a Stratasys Objet Connex 500 3D printer. The mold contains a pattern of $5 \times 5$ circular holes in which we place steel rods. The holes have a diameter of 10mm and wall thickness of 1.5 mm, implying a sample width and height of 59 mm. The depth of the sample is 50 mm. The sample is connected to two perspex plates using silicone rubber glue. The perspex plates can be used to mount the sample to a test rig at the bottom and to mount additional weight and an accelerometer at the top.

\subsubsection{Multi-directional buckling sample fabrication}
\label{subsec:MultiFabri}
The sample of Fig. \ref{fig:bidir}A-D has been manufactured from a 2-component slow curing silicone rubber using an EnvisionTEC 3D Bioplotter. The sample measures 240 $\times$ 115 $\times$ 22 mm.

\subsection{Experimental methods}

\subsubsection{Uniaxial testing}
\label{subsec:app_uniaxial}
To perform the tests of Fig. \ref{fig2}AB and Fig. \ref{fig:metal}B, we compress and extend the samples using a uniaxial testing device (Instron 5943 with a 500 N load cell). We do so at constant rates of 1 mm/min (Fig. \ref{fig:metal}B) and 1, 10, 100 and 1000 mm/min (Fig. \ref{fig2}AB). 

\subsubsection{Vibration testing}
\label{sec:vibtest}
To perform the vibration tests of Fig. \ref{fig1}, Fig. \ref{fig2} and Fig. \ref{fig:random}, we vibrate the sample using an Instron Electropuls 3000. We suspend a rectangular steel frame from the machine and mount the sample on the bottom of this frame, such that we effectively obtain a base exciting from the bottom. A mass is mounted on top of the frame to effectively obtain a mass-spring system with base excitation (Fig. \ref{fig1}CD. 

\begin{figure}
\centering 
\includegraphics[width=.5\columnwidth]{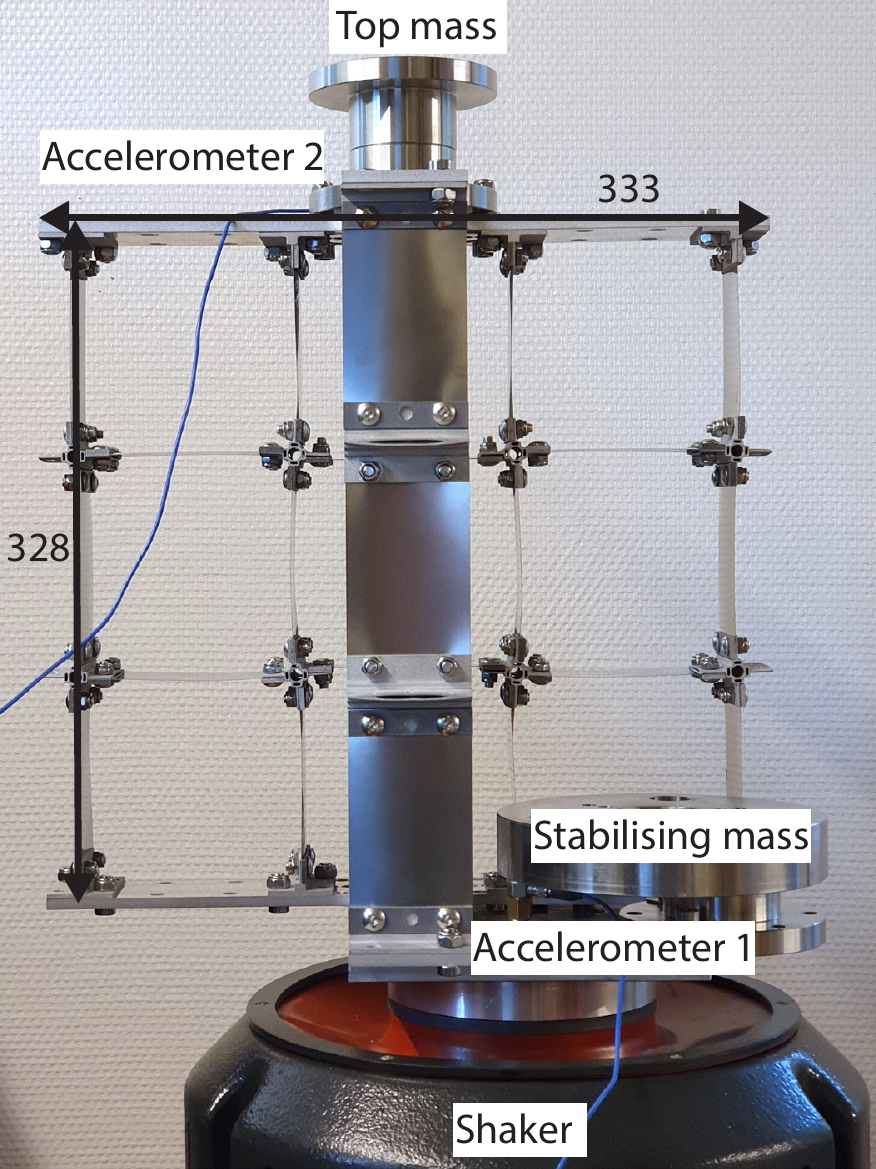}
\caption{Metal buckling based vibration damping sample mounted on shaker table.}
\label{fig:MetalMethods}
\end{figure}

For the vibration tests with the metal sample of Fig. \ref{fig:metal}, we vibrate the sample using a Tira Vibration Test System TV 5220-120 instead, where we control the frequency with an Aim-TTi TG5011 function generator. We add a large mass of 4.5 kg on the shaker next to the sample to keep the vibration input acceleration level more constant during a frequency sweep. The set-up is shown in Fig. \ref{fig:MetalMethods}.

In both cases, we measure the accelerations at 5000 Hz with two PCB Piezotronics 352C33 accelerometers: one at the bottom, capturing the base excitation and one on top, capturing the output accelerations. The tests of Fig. \ref{fig1}, Fig. \ref{fig2} are also recorded using a Phantom VEO 640 monochrome high-speed CMOS camera with Nikon 200mm f/4 Macro lens, at 250 frames per second at a resolution of $1024 \times 1024$, inducing a spatial resolution of $0.10$ mm.  

\subsection{Processing accelerometer data}
To process the accelerometer data, we first apply a fourth order Butterworth bandpass filter from 4 to 60 Hz for the elastomeric sample and from 5 to 100 Hz for the metal sample. This way, we filter out low frequency sensor drift and high frequency noise generated by the test set-up. 

We then use the Python \textit{scipy.signal.find\_peaks} algorithm to find the trends of the peak acceleration and as function of time and frequency. We also apply a Savitzky-Golay filter to smooth the resulting peak acceleration trends.

\subsection{Image analysis}
\label{sec:image}
To get a quantitative understanding of the degree of buckling of the sample used in Fig. \ref{fig1} and Fig. \ref{fig2}, we use particle tracking (OPENCV and Python) and custom-made tracking algorithms to quantify the flattening $f$ and orientation $\phi$ w.r.t. the horizontal of each pore and calculate the polarisation~\cite{florijn2014programmable,dykstra2019viscoelastic,bossart2021oligomodal}:

\begin{equation}
\Omega_{n_x n_y}:=(-1)^{n_x+n_y} f\cos 2\phi ,    
\end{equation} 

where $n_x$ ($n_y$) is the hole's column (row). We then plot these tracked ellipses along with their polarisation in red and blue in Fig. \ref{fig1}CD. 

\subsection{Numerical model}
\label{sec:app_num}
The numerical model consists of a nonlinear mass spring damper system (Fig. \ref{fig1}A). The system contains a mass $M$, output displacement $y(t)$ and acceleration $\dot{\dot{y}}$ (orange in Fig. \ref{fig1}A),coupled through a spring to a base excitation $u(t)$, $\dot{\dot{u}}$ (blue in Fig. \ref{fig1}A). The spring extension $s = y-u$. The spring stiffness $K$ is multilinear, while the damper $C$ is multilinear only in the spring extension $s$. $K$ and $C$ can be normalized with the mass to obtain the $k$ and $c$. These can then be coupled to the linear system response as follows, where $\omega_n = \frac{f_n}{2 \pi}$ is the angular eigenfrequency, $\zeta$ is the damping coefficient, $G$ is the amplification factor and $f_n$ is the eigenfrequency: 

\begin{equation}
    k = \frac{K}{m} =  \omega_n^2
\end{equation}

\begin{equation}
    c = \frac{C}{m} = 2\zeta \omega_n
\end{equation}

\begin{equation}
    \zeta = \frac{1}{2G}
\end{equation}

We derive the linearized spring stiffness $K$, eigenfrequency $f_n$ and amplification factor $G$ from the experimental data of Fig. \ref{fig2}A,C for the elastomeric sample and Fig. \ref{fig:metal}B as well as the underlying data at small excitations of Fig. \ref{fig:metal}D for the metal sample. We derive for small excitations, $K = 11.38$ N/mm, $f_n = 32$ and $G=19$ for the elastomeric sample and $K = 368.8$ N/mm, $f_n = 48$ Hz and $G=13.2$ for the metal sample. We use these values to derive $M=0.28$ kg and $C=2.83$ Ns/m for the elastomeric sample and $M=4.0$ kg and $C=92.6$ Ns/m for the metal sample. These masses are higher than the actual masses mounted on top as the sample mass also affects the eigenfrequency. We also derive the buckling force plateaus to be 25 N and 78 N, for the elastomeric (Fig. \ref{fig2}E) and metal (Fig. \ref{fig:metal}) samples respectively. To account for gravity, we subtract the gravitational acceleration, $g M$, where $g=9.81$ m/s$^2$ in the model to end up with the base state at $s=0$ without excitation in the model. The force-displacement curve for the model with buckling in two directions, seen in Fig. \ref{fig:bidir}E presents the same force plateau as the one for the metal sample in Fig. \ref{fig:metal}G, with $g M$ subtracted.  

The system is governed by the following differential equation.

\begin{equation}
    \dot{\dot{y}} +c\left(\dot{y}-\dot{u} \right)  +k\left(y-u \right) = 0
    \label{eq:dif}
\end{equation}

When the system is subjected to a base excitation $u(t)$, $\dot{\dot{u}}$, the system can be solved numerically in the time domain using a forward Taylor series, which we do at a frequency of 2000 Hz:

\begin{equation}
    \dot{\dot{y}} = -c\left(\dot{y}-\dot{u} \right)  -k\left(y-u \right) 
\end{equation}

\begin{equation}
    \dot{\dot{\dot{y}}} = -c\left(\, \dot{\dot{y}} -\dot{\dot{u}} \right)  -k\left(\dot{y}-\dot{u} \right) 
\end{equation}

Compared to some alternative ways to numerically solve the system, a forward Taylor series has a number of distinct advantages. Compared to closed form solutions, such as piecewise solutions~\cite{shaw1983periodically,jung2014nonlinear}, a forward Taylor series has the distinct advantage that it can easily be modified to allow for more complicated stiffness and damping characteristics. Compared to Euler's method, which is a first order Taylor series, a higher order forward Taylor series offers more numerical stability and offers a much higher speed of convergence. Finally, compared to explicit nonlinear finite element methods, a simple single degree of freedom system offers large advantages in simplicity, computational needs and numerical stability, at the cost of a lower accuracy.

\subsection{Defining the vibration signal}
\label{subsec:app_defining}
To obtain the frequency response, we subject the numerical and experimental sample to a sinusoidal vibration with constant acceleration amplitude, where the sine slightly changes frequency after every cycle. At each start of a full cycle, implying the acceleration and displacement are 0, the frequency of the sine cycle is updated. This implies that the input acceleration, velocity and displacement are defined as follow, with $B$ the input acceleration amplitude, $f$ the frequency and $t_c$ the time in the cycle between $t_c=0$s to $t_c=\frac{1}{f}$:

\begin{equation}
    \dot{\dot{u}} = B \sin{\left(2 \pi f t_c\right)} 
\end{equation}

\begin{equation}
    \dot{u} = -\frac{B}{2 \pi f} \cos{\left(2 \pi f t_c\right)}
\end{equation}

\begin{equation}
    u = -\frac{B}{\left(2 \pi f\right)^2} \sin{\left(2 \pi f t_c\right)}
\end{equation}

\subsection{Random vibration}
\label{app_random}

\begin{figure}[ht!]
\centering 
\includegraphics[width=\columnwidth]{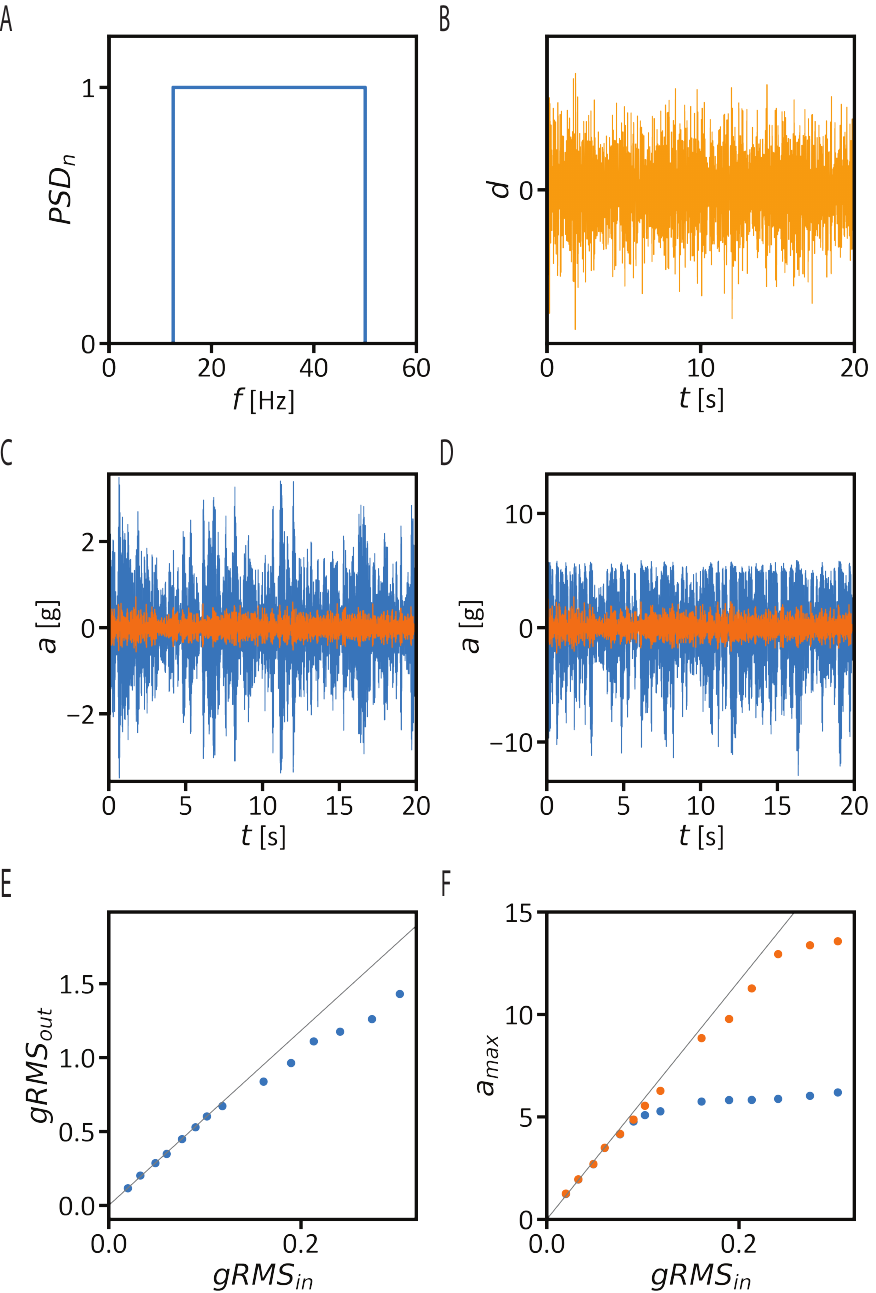}
\caption{\textbf{Random vibration.} (A) Power Spectral Density (PSD) profile of the input acceleration, normalized with the peak PSD level. (B) Corresponding time-displacement history applied in the tests. (C,D) Experimental time-acceleration history for input accelerations of 0.06 gRMS(c) and 0.30 gRMS(d). Orange and blue curves correspond to the input and out accelerations respectively. (E) Output acceleration (gRMS$_{out}$) as function of input acceleration (gRMS$_{in}$). (F) Maximum acceleration, $a_{max}$, as function of the input acceleration. Blue and orange correspond to compression and tension respectively. All RMS values are shown at 1$\sigma$.}
\label{fig:random}
\end{figure}

We first define a random vibration spectrum. We start from the acceleration Power Spectral Density Function of Fig. \ref{fig:random}A, calculated at 1 standard deviation (1$\sigma$). $PSD_n$ is the PSD value, normalized with the maximum PSD value. This PSD shows a constant PSD value between 12.5 Hz and 50 Hz, which also includes the area surrounding the eigenfrequency of the sample of Fig. \ref{fig1}.  This implies that vibration with this PSD will only excite the frequencies around the eigenfrequency. 

The 1$\sigma$ root-mean-square of any PSD spectrum can be calculated as~\cite{irvine2009introduction}:

\begin{equation}
    RMS = \sqrt{\int_0^\infty PSD\, d f}. 
\end{equation}

We can convert the acceleration PSD to a time signal using an inverse Fourier transform~\cite{irvine2009introduction}, where we assume random values for the starting phase of the equivalent wave at each frequency. We can integrate the time-acceleration signal to a time-velocity and time-displacement signal and use a high-pass Butterworth filter to remove low-frequency drift. As such, we can obtain the time-displacement signal of Fig. \ref{fig1}B. We can also convert time signals to PSD signals using a Fourier transform.

We can then shake the sample, as described in the Main Text and in Section~\ref{sec:vibtest}, using this time signal as an input. If we subject the sample to a low random base excitation of 0.06 gRMS in Fig. \ref{fig:random}D, we observe that the output acceleration in blue is much larger than the input acceleration in orange and that acceleration levels are similar in tension and compression. However, when we do the same at a higher base excitation of 0.30 gRMS in Fig. \ref{fig:random}D, we observe a clear upper limit in compression, due to buckling.

We can also track how the output accelerations change with the applied input acceleration. 
In Fig. \ref{fig:random}E, we observe for $gRMS_{in} > 0.15$ that $gRMS_{out}$  starts to go lower than the linear trend, which suggests that buckling based vibration damping also works in random vibrations. However, to convert a time signal to a frequency based signal, including a gRMS value, the implicit assumption is made that the signal is a sum of sinusoidal signals, which is approximately symmetric about 0. That is not the case here, as seen in Fig. \ref{fig:random}D. As such, it is more representative to look at the peak accelerations instead. These are plotted in Fig. \ref{fig:random}F. Here we observe a clear upper limit in compression (blue) for  $gRMS_{in} > 0.1$. However, for $gRMS_{in} > 0.2$, we observe a similar plateau in tension (orange). We presume that the reason why it requires a much larger base excitation to set an upper limit in tension, is because only a few cycles induce buckling and therefore induce additional dissipation. This is different from vibrations at constant frequency and acceleration, where buckling occurs every cycle, along with the accompanying increased dissipation. However, both Fig. \ref{fig:random}E and Fig. \ref{fig:random}F demonstrate that buckling based vibration damping also works in tension.

\subsection{Design \& Finite Element Analysis}
\label{subsec:FEM}

\begin{figure*}[ht!]
\centering 
\includegraphics[width=1.\textwidth]{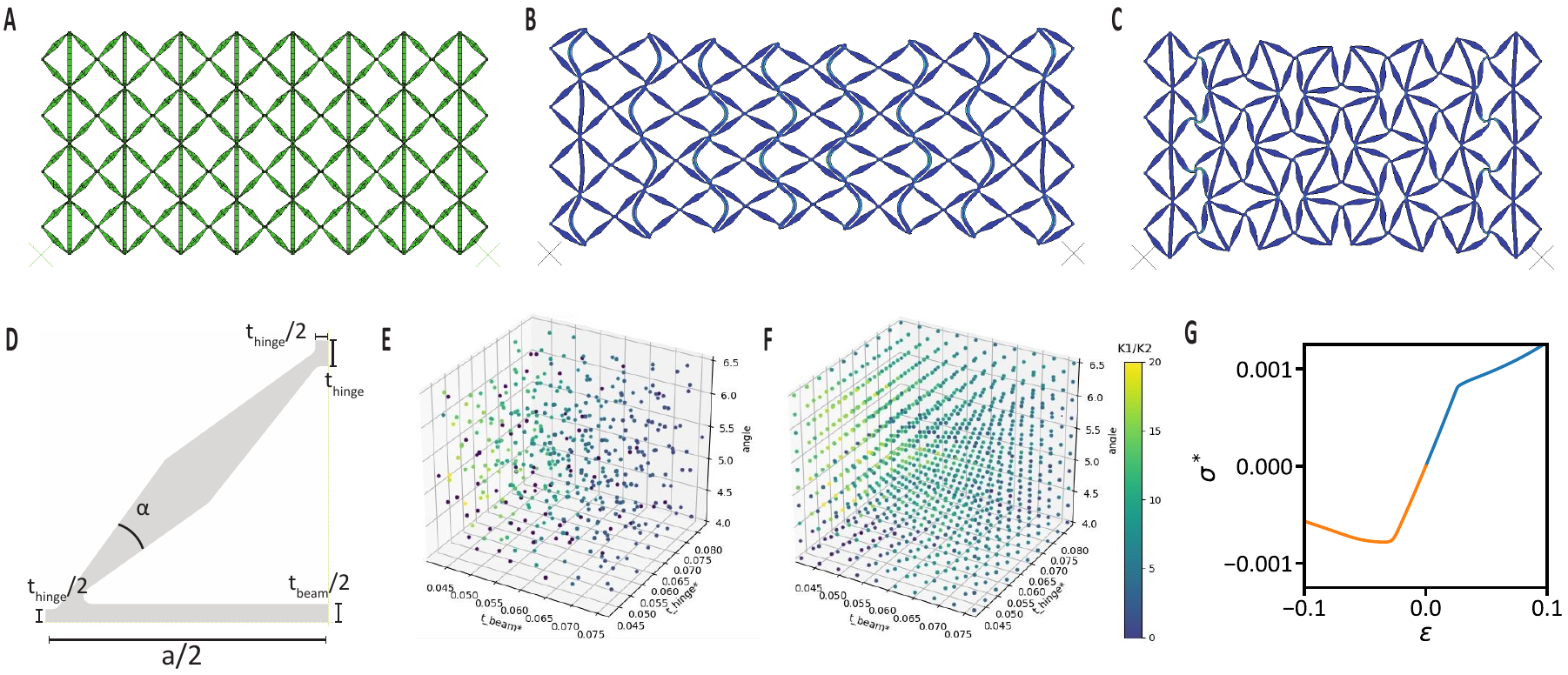}
\caption{\textbf{Simulating and optimising bi-directional buckling metamaterials}. (A) Finite element mesh. (B) Tensile response. (C) Compressive response. (D) Quarter unit cell. (E) Stiffness ratios for training points before and after buckling: minimum of tension of compression. (F) Stiffness ratios interpolated with Gaussian Processes Regression before and after buckling: minimum of tension of compression. (G) Normalized stress ($\sigma^*$)-strain ($\epsilon)$ curve in tension (blue) and compression (orange) of sample in A.}
\label{fig:bidir_appendix}
\end{figure*}
For the sample of Fig. \ref{fig:bidir}, we designed the sample using a form of Bayesian optimisation combined with finite element simulations in Abaqus~\cite{bessa2019bayesian}.

To predict the response of the metamaterial samples of Fig. \ref{fig:metal} and Fig. \ref{fig:bidir}, we use nonlinear finite element simulations in Abaqus 2021 (Dassault Systèmes). 

\subsubsection{Metal sample}
We designed the sample of Fig. \ref{fig:metal} using a combination of finite element analyses and experiments. We started designing the sample using 3D analyses with 2D plane stress plate elements (CPS8R). We used a combination of single strip analyses, $3\times 3$ and $5\times 5$ metamaterial analyses. We modelled the entire sample using aluminium with a Young's modulus of 70 GPa, Poisson's ratio of 0.3 and density of 2800 kg/m$^3$. We performed various sets of analyses, varying the aspect ratio, thickness and out-of-plane curvature. We compressed and uncompressed the sample numerically using a nonlinear quasi-static analysis. In some cases, snap-through instabilities made the analyses unstable, in which cases we opted for dynamic explicit analyses instead. We sized the design using two main criteria: 
\begin{enumerate}
    \item At fives times the post-buckling strain, the force-drop from peak is less than 50\% and a $5\times 5$ aluminium sample still has at least a 40 N force. This prevents unstable collapse post-buckling.
    \item At fives times the post-buckling strain, the maximum Von Mises Stress is less than 100 MPa in aluminium. This prevents fatigue damage in high strength metals. 
\end{enumerate}
After analysis, we opted to make the sample out of steel sheets instead due to the higher load carrying capacity and better availability of high-strength variants with small thicknesses (AISI 301 Full Hard). After producing the first sample, we opted to produce the final sample of Fig. \ref{fig:metal} out of a higher thickness than originally analysed as the produced load carrying capacity was lower than predicted due to imperfections.

\subsubsection{Bi-directional buckling sample}

The final design and mesh of the sample of Fig. \ref{fig:bidir} is given in Fig. \ref{fig:bidir_appendix}A. We modelled the sample in 2D using quadratic quad-dominated CPE8R plane strain elements. For the material, we assumed a Neo-Hookean material model with a Poisson's of 0.48. We fully constrained the left and right side of Fig. \ref{fig:bidir_appendix}A and compressed and extended the sample numerically using a nonlinear quasi-static analysis.

To maximise the efficiency of bi-directional buckling based vibration damping, the sample had to buckle in both directions at an equal strain. To obtain a suitable design, we used Bayesian optimisation~\cite{masurier2021thesis,bessa2019bayesian}. First, we specified the design using the parameters as shown in the quarter unit cell of Fig. \ref{fig:bidir_appendix}D. We optimized the design for three normalized design parameters: $t_{beam}^{*} = \frac{t_{beam}}{a}$, $t_{hinge}^{*} = \frac{t_{hinge}}{a}$ and $angle = \alpha$. We ran 500 analyses with a variety of combinations of these three variables. We populated the design space using a Sobol sequence, an efficient method to populate a multivariable design space with an arbitrary number of data points~\cite{sobol1967distribution}. We tracked two outputs: (i) the ratio in stiffness before and after buckling, $K_1/K_2$: minimum of tension and compression, and (ii) the ratio of the buckling force in tension and compressive direction, $F_{b,t} / F_{b,c}$. For $K_1/K_2$, we plotted the results in Fig. \ref{fig:bidir_appendix}E. We then used Bayesian machine learning (Gaussian Processes Regression) to more densely interpolate the design space as seen in Fig. \ref{fig:bidir_appendix}F for $K_1 / K_2$. In doing so, we also obtained the standard deviation, $\sigma$, of the uncertainty of the interpolation. We then defined the following criteria to define designs of sufficient quality:

\begin{enumerate}
    \item $K_1/K_2 > 5 +1.96 \sigma_{K_1/K_2}$
    \item $|F_{b,t} / F_{b,c}|-1 < 0.15- 1.96\sigma_{F_{b,t} / F_{b,c}}$
\end{enumerate}

This presents us with several designs which adhere to our requirements. Around these locations, we then refined our data by running 100 additional analyses using a Sobol sequence, which then accurately provided us with a variety of designs adhering to our requirements. One of those is the design of Fig. \ref{fig:bidir_appendix}A-C, which we produced in Fig. \ref{fig:bidir}. The nonlinear normalized stress-strain curve as calculated by Abaqus, is presented in Fig. \ref{fig:bidir_appendix}G, where the stress has been normalized by the cross-section and Young's modulus.

\section{Data and Code Availability}
The data and codes that support the figures within this paper are publicly available on a Zenodo repository~\cite{Zenodo2023}. Two videos that support this article can also be found in the supplementary information. Supplementary Videos 1 and 2 show buckling of samples with base excitations around resonance for the rubber sample of Fig. \ref{fig1} and for the metal sample of Fig. \ref{fig:metal} respectively.

\section{Acknowledgements}
We thank Jackson Wilt for helping us with the manufacturing of of the bi-directional buckling sample, Daan Giessen, Sven Koot, Clint Ederveen Janssen and Taco Walstra for technical support and Marc Serra Garcia for reviewing our article. We acknowledge funding from the European Research Council under grant agreement 852587 and the Netherlands Organisation for Scientific Research under grant agreement NWO TTW 17883.


\bibliography{references}

\providecommand{\noopsort}[1]{}\providecommand{\singleletter}[1]{#1}%
\begin{thebibliography}{52}%
\makeatletter
\providecommand \@ifxundefined [1]{%
 \@ifx{#1\undefined}
}%
\providecommand \@ifnum [1]{%
 \ifnum #1\expandafter \@firstoftwo
 \else \expandafter \@secondoftwo
 \fi
}%
\providecommand \@ifx [1]{%
 \ifx #1\expandafter \@firstoftwo
 \else \expandafter \@secondoftwo
 \fi
}%
\providecommand \natexlab [1]{#1}%
\providecommand \enquote  [1]{``#1''}%
\providecommand \bibnamefont  [1]{#1}%
\providecommand \bibfnamefont [1]{#1}%
\providecommand \citenamefont [1]{#1}%
\providecommand \href@noop [0]{\@secondoftwo}%
\providecommand \href [0]{\begingroup \@sanitize@url \@href}%
\providecommand \@href[1]{\@@startlink{#1}\@@href}%
\providecommand \@@href[1]{\endgroup#1\@@endlink}%
\providecommand \@sanitize@url [0]{\catcode `\\12\catcode `\$12\catcode
  `\&12\catcode `\#12\catcode `\^12\catcode `\_12\catcode `\%12\relax}%
\providecommand \@@startlink[1]{}%
\providecommand \@@endlink[0]{}%
\providecommand \url  [0]{\begingroup\@sanitize@url \@url }%
\providecommand \@url [1]{\endgroup\@href {#1}{\urlprefix }}%
\providecommand \urlprefix  [0]{URL }%
\providecommand \Eprint [0]{\href }%
\providecommand \doibase [0]{https://doi.org/}%
\providecommand \selectlanguage [0]{\@gobble}%
\providecommand \bibinfo  [0]{\@secondoftwo}%
\providecommand \bibfield  [0]{\@secondoftwo}%
\providecommand \translation [1]{[#1]}%
\providecommand \BibitemOpen [0]{}%
\providecommand \bibitemStop [0]{}%
\providecommand \bibitemNoStop [0]{.\EOS\space}%
\providecommand \EOS [0]{\spacefactor3000\relax}%
\providecommand \BibitemShut  [1]{\csname bibitem#1\endcsname}%
\let\auto@bib@innerbib\@empty
\bibitem [{\citenamefont {Zhou}\ \emph {et~al.}(2016)\citenamefont {Zhou},
  \citenamefont {Yu}, \citenamefont {Shao}, \citenamefont {Zhang},\ and\
  \citenamefont {Wang}}]{zhou2016research}%
  \BibitemOpen
  \bibfield  {author} {\bibinfo {author} {\bibfnamefont {X.}~\bibnamefont
  {Zhou}}, \bibinfo {author} {\bibfnamefont {D.}~\bibnamefont {Yu}}, \bibinfo
  {author} {\bibfnamefont {X.}~\bibnamefont {Shao}}, \bibinfo {author}
  {\bibfnamefont {S.}~\bibnamefont {Zhang}},\ and\ \bibinfo {author}
  {\bibfnamefont {S.}~\bibnamefont {Wang}},\ }\bibfield  {title} {\bibinfo
  {title} {Research and applications of viscoelastic vibration damping
  materials: A review},\ }\href@noop {} {\bibfield  {journal} {\bibinfo
  {journal} {Composite Structures}\ }\textbf {\bibinfo {volume} {136}},\
  \bibinfo {pages} {460} (\bibinfo {year} {2016})}\BibitemShut {NoStop}%
\bibitem [{\citenamefont {Baz}(2019)}]{baz2019active}%
  \BibitemOpen
  \bibfield  {author} {\bibinfo {author} {\bibfnamefont {A.~M.}\ \bibnamefont
  {Baz}},\ }\href@noop {} {\emph {\bibinfo {title} {Active and passive
  vibration damping}}}\ (\bibinfo  {publisher} {John Wiley \& Sons},\ \bibinfo
  {year} {2019})\BibitemShut {NoStop}%
\bibitem [{\citenamefont {Lakes}\ \emph {et~al.}(2001)\citenamefont {Lakes},
  \citenamefont {Lee}, \citenamefont {Bersie},\ and\ \citenamefont
  {Wang}}]{lakes2001extreme}%
  \BibitemOpen
  \bibfield  {author} {\bibinfo {author} {\bibfnamefont {R.~S.}\ \bibnamefont
  {Lakes}}, \bibinfo {author} {\bibfnamefont {T.}~\bibnamefont {Lee}}, \bibinfo
  {author} {\bibfnamefont {A.}~\bibnamefont {Bersie}},\ and\ \bibinfo {author}
  {\bibfnamefont {Y.-C.}\ \bibnamefont {Wang}},\ }\bibfield  {title} {\bibinfo
  {title} {Extreme damping in composite materials with negative-stiffness
  inclusions},\ }\href@noop {} {\bibfield  {journal} {\bibinfo  {journal}
  {Nature}\ }\textbf {\bibinfo {volume} {410}},\ \bibinfo {pages} {565}
  (\bibinfo {year} {2001})}\BibitemShut {NoStop}%
\bibitem [{\citenamefont {Wang}\ and\ \citenamefont
  {Lakes}(2004)}]{wang2004extreme}%
  \BibitemOpen
  \bibfield  {author} {\bibinfo {author} {\bibfnamefont {Y.}~\bibnamefont
  {Wang}}\ and\ \bibinfo {author} {\bibfnamefont {R.}~\bibnamefont {Lakes}},\
  }\bibfield  {title} {\bibinfo {title} {Extreme stiffness systems due to
  negative stiffness elements},\ }\href@noop {} {\bibfield  {journal} {\bibinfo
   {journal} {American Journal of Physics}\ }\textbf {\bibinfo {volume} {72}},\
  \bibinfo {pages} {40} (\bibinfo {year} {2004})}\BibitemShut {NoStop}%
\bibitem [{\citenamefont {Jaglinski}\ \emph {et~al.}(2007)\citenamefont
  {Jaglinski}, \citenamefont {Kochmann}, \citenamefont {Stone},\ and\
  \citenamefont {Lakes}}]{jaglinski2007composite}%
  \BibitemOpen
  \bibfield  {author} {\bibinfo {author} {\bibfnamefont {T.}~\bibnamefont
  {Jaglinski}}, \bibinfo {author} {\bibfnamefont {D.}~\bibnamefont {Kochmann}},
  \bibinfo {author} {\bibfnamefont {D.}~\bibnamefont {Stone}},\ and\ \bibinfo
  {author} {\bibfnamefont {R.~S.}\ \bibnamefont {Lakes}},\ }\bibfield  {title}
  {\bibinfo {title} {Composite materials with viscoelastic stiffness greater
  than diamond},\ }\href@noop {} {\bibfield  {journal} {\bibinfo  {journal}
  {Science}\ }\textbf {\bibinfo {volume} {315}},\ \bibinfo {pages} {620}
  (\bibinfo {year} {2007})}\BibitemShut {NoStop}%
\bibitem [{\citenamefont {Kochmann}\ and\ \citenamefont
  {Bertoldi}(2017)}]{kochmann2017exploiting}%
  \BibitemOpen
  \bibfield  {author} {\bibinfo {author} {\bibfnamefont {D.~M.}\ \bibnamefont
  {Kochmann}}\ and\ \bibinfo {author} {\bibfnamefont {K.}~\bibnamefont
  {Bertoldi}},\ }\bibfield  {title} {\bibinfo {title} {Exploiting
  microstructural instabilities in solids and structures: from metamaterials to
  structural transitions},\ }\href@noop {} {\bibfield  {journal} {\bibinfo
  {journal} {Applied mechanics reviews}\ }\textbf {\bibinfo {volume} {69}}
  (\bibinfo {year} {2017})}\BibitemShut {NoStop}%
\bibitem [{\citenamefont {Li}\ \emph {et~al.}(2020)\citenamefont {Li},
  \citenamefont {Li},\ and\ \citenamefont {Li}}]{li2020negative}%
  \BibitemOpen
  \bibfield  {author} {\bibinfo {author} {\bibfnamefont {H.}~\bibnamefont
  {Li}}, \bibinfo {author} {\bibfnamefont {Y.}~\bibnamefont {Li}},\ and\
  \bibinfo {author} {\bibfnamefont {J.}~\bibnamefont {Li}},\ }\bibfield
  {title} {\bibinfo {title} {Negative stiffness devices for vibration isolation
  applications: A review},\ }\href@noop {} {\bibfield  {journal} {\bibinfo
  {journal} {Advances in Structural Engineering}\ }\textbf {\bibinfo {volume}
  {23}},\ \bibinfo {pages} {1739} (\bibinfo {year} {2020})}\BibitemShut
  {NoStop}%
\bibitem [{\citenamefont {Balaji}\ and\ \citenamefont
  {Karthik~SelvaKumar}(2021)}]{balaji2021applications}%
  \BibitemOpen
  \bibfield  {author} {\bibinfo {author} {\bibfnamefont {P.}~\bibnamefont
  {Balaji}}\ and\ \bibinfo {author} {\bibfnamefont {K.}~\bibnamefont
  {Karthik~SelvaKumar}},\ }\bibfield  {title} {\bibinfo {title} {Applications
  of nonlinearity in passive vibration control: a review},\ }\href@noop {}
  {\bibfield  {journal} {\bibinfo  {journal} {Journal of Vibration Engineering
  \& Technologies}\ }\textbf {\bibinfo {volume} {9}},\ \bibinfo {pages} {183}
  (\bibinfo {year} {2021})}\BibitemShut {NoStop}%
\bibitem [{\citenamefont {Lu}\ \emph {et~al.}(2009)\citenamefont {Lu},
  \citenamefont {Feng},\ and\ \citenamefont {Chen}}]{lu2009phononic}%
  \BibitemOpen
  \bibfield  {author} {\bibinfo {author} {\bibfnamefont {M.-H.}\ \bibnamefont
  {Lu}}, \bibinfo {author} {\bibfnamefont {L.}~\bibnamefont {Feng}},\ and\
  \bibinfo {author} {\bibfnamefont {Y.-F.}\ \bibnamefont {Chen}},\ }\bibfield
  {title} {\bibinfo {title} {Phononic crystals and acoustic metamaterials},\
  }\href@noop {} {\bibfield  {journal} {\bibinfo  {journal} {Materials today}\
  }\textbf {\bibinfo {volume} {12}},\ \bibinfo {pages} {34} (\bibinfo {year}
  {2009})}\BibitemShut {NoStop}%
\bibitem [{\citenamefont {Hussein}\ and\ \citenamefont
  {Frazier}(2010)}]{hussein2010band}%
  \BibitemOpen
  \bibfield  {author} {\bibinfo {author} {\bibfnamefont {M.~I.}\ \bibnamefont
  {Hussein}}\ and\ \bibinfo {author} {\bibfnamefont {M.~J.}\ \bibnamefont
  {Frazier}},\ }\bibfield  {title} {\bibinfo {title} {Band structure of
  phononic crystals with general damping},\ }\href@noop {} {\bibfield
  {journal} {\bibinfo  {journal} {Journal of Applied Physics}\ }\textbf
  {\bibinfo {volume} {108}},\ \bibinfo {pages} {093506} (\bibinfo {year}
  {2010})}\BibitemShut {NoStop}%
\bibitem [{\citenamefont {Hussein}\ and\ \citenamefont
  {Frazier}(2013)}]{hussein2013metadamping}%
  \BibitemOpen
  \bibfield  {author} {\bibinfo {author} {\bibfnamefont {M.~I.}\ \bibnamefont
  {Hussein}}\ and\ \bibinfo {author} {\bibfnamefont {M.~J.}\ \bibnamefont
  {Frazier}},\ }\bibfield  {title} {\bibinfo {title} {Metadamping: An emergent
  phenomenon in dissipative metamaterials},\ }\href@noop {} {\bibfield
  {journal} {\bibinfo  {journal} {Journal of Sound and Vibration}\ }\textbf
  {\bibinfo {volume} {332}},\ \bibinfo {pages} {4767} (\bibinfo {year}
  {2013})}\BibitemShut {NoStop}%
\bibitem [{\citenamefont {Wang}\ \emph {et~al.}(2014)\citenamefont {Wang},
  \citenamefont {Casadei}, \citenamefont {Shan}, \citenamefont {Weaver},\ and\
  \citenamefont {Bertoldi}}]{wang2014harnessing}%
  \BibitemOpen
  \bibfield  {author} {\bibinfo {author} {\bibfnamefont {P.}~\bibnamefont
  {Wang}}, \bibinfo {author} {\bibfnamefont {F.}~\bibnamefont {Casadei}},
  \bibinfo {author} {\bibfnamefont {S.}~\bibnamefont {Shan}}, \bibinfo {author}
  {\bibfnamefont {J.~C.}\ \bibnamefont {Weaver}},\ and\ \bibinfo {author}
  {\bibfnamefont {K.}~\bibnamefont {Bertoldi}},\ }\bibfield  {title} {\bibinfo
  {title} {Harnessing buckling to design tunable locally resonant acoustic
  metamaterials},\ }\href@noop {} {\bibfield  {journal} {\bibinfo  {journal}
  {Physical review letters}\ }\textbf {\bibinfo {volume} {113}},\ \bibinfo
  {pages} {014301} (\bibinfo {year} {2014})}\BibitemShut {NoStop}%
\bibitem [{\citenamefont {Cummer}\ \emph {et~al.}(2016)\citenamefont {Cummer},
  \citenamefont {Christensen},\ and\ \citenamefont
  {Al{\`u}}}]{cummer2016controlling}%
  \BibitemOpen
  \bibfield  {author} {\bibinfo {author} {\bibfnamefont {S.~A.}\ \bibnamefont
  {Cummer}}, \bibinfo {author} {\bibfnamefont {J.}~\bibnamefont
  {Christensen}},\ and\ \bibinfo {author} {\bibfnamefont {A.}~\bibnamefont
  {Al{\`u}}},\ }\bibfield  {title} {\bibinfo {title} {Controlling sound with
  acoustic metamaterials},\ }\href@noop {} {\bibfield  {journal} {\bibinfo
  {journal} {Nature Reviews Materials}\ }\textbf {\bibinfo {volume} {1}},\
  \bibinfo {pages} {1} (\bibinfo {year} {2016})}\BibitemShut {NoStop}%
\bibitem [{\citenamefont {Krushynska}\ \emph {et~al.}(2017)\citenamefont
  {Krushynska}, \citenamefont {Miniaci}, \citenamefont {Bosia},\ and\
  \citenamefont {Pugno}}]{krushynska2017coupling}%
  \BibitemOpen
  \bibfield  {author} {\bibinfo {author} {\bibfnamefont {A.~O.}\ \bibnamefont
  {Krushynska}}, \bibinfo {author} {\bibfnamefont {M.}~\bibnamefont {Miniaci}},
  \bibinfo {author} {\bibfnamefont {F.}~\bibnamefont {Bosia}},\ and\ \bibinfo
  {author} {\bibfnamefont {N.~M.}\ \bibnamefont {Pugno}},\ }\bibfield  {title}
  {\bibinfo {title} {Coupling local resonance with bragg band gaps in
  single-phase mechanical metamaterials},\ }\href@noop {} {\bibfield  {journal}
  {\bibinfo  {journal} {Extreme Mechanics Letters}\ }\textbf {\bibinfo {volume}
  {12}},\ \bibinfo {pages} {30} (\bibinfo {year} {2017})}\BibitemShut {NoStop}%
\bibitem [{\citenamefont {Gao}\ \emph {et~al.}(2022)\citenamefont {Gao},
  \citenamefont {Zhang}, \citenamefont {Deng}, \citenamefont {Guo},
  \citenamefont {Cheng},\ and\ \citenamefont {Hou}}]{gao2022acoustic}%
  \BibitemOpen
  \bibfield  {author} {\bibinfo {author} {\bibfnamefont {N.}~\bibnamefont
  {Gao}}, \bibinfo {author} {\bibfnamefont {Z.}~\bibnamefont {Zhang}}, \bibinfo
  {author} {\bibfnamefont {J.}~\bibnamefont {Deng}}, \bibinfo {author}
  {\bibfnamefont {X.}~\bibnamefont {Guo}}, \bibinfo {author} {\bibfnamefont
  {B.}~\bibnamefont {Cheng}},\ and\ \bibinfo {author} {\bibfnamefont
  {H.}~\bibnamefont {Hou}},\ }\bibfield  {title} {\bibinfo {title} {Acoustic
  metamaterials for noise reduction: a review},\ }\href@noop {} {\bibfield
  {journal} {\bibinfo  {journal} {Advanced Materials Technologies}\ ,\ \bibinfo
  {pages} {2100698}} (\bibinfo {year} {2022})}\BibitemShut {NoStop}%
\bibitem [{\citenamefont {Collette}\ \emph {et~al.}(2011)\citenamefont
  {Collette}, \citenamefont {Janssens},\ and\ \citenamefont
  {Artoos}}]{collette2011review}%
  \BibitemOpen
  \bibfield  {author} {\bibinfo {author} {\bibfnamefont {C.}~\bibnamefont
  {Collette}}, \bibinfo {author} {\bibfnamefont {S.}~\bibnamefont {Janssens}},\
  and\ \bibinfo {author} {\bibfnamefont {K.}~\bibnamefont {Artoos}},\
  }\bibfield  {title} {\bibinfo {title} {Review of active vibration isolation
  strategies},\ }\href@noop {} {\bibfield  {journal} {\bibinfo  {journal}
  {Recent patents on Mechanical engineering}\ }\textbf {\bibinfo {volume}
  {4}},\ \bibinfo {pages} {212} (\bibinfo {year} {2011})}\BibitemShut {NoStop}%
\bibitem [{\citenamefont {Aridogan}\ and\ \citenamefont
  {Basdogan}(2015)}]{aridogan2015review}%
  \BibitemOpen
  \bibfield  {author} {\bibinfo {author} {\bibfnamefont {U.}~\bibnamefont
  {Aridogan}}\ and\ \bibinfo {author} {\bibfnamefont {I.}~\bibnamefont
  {Basdogan}},\ }\bibfield  {title} {\bibinfo {title} {A review of active
  vibration and noise suppression of plate-like structures with piezoelectric
  transducers},\ }\href@noop {} {\bibfield  {journal} {\bibinfo  {journal}
  {Journal of Intelligent Material Systems and Structures}\ }\textbf {\bibinfo
  {volume} {26}},\ \bibinfo {pages} {1455} (\bibinfo {year}
  {2015})}\BibitemShut {NoStop}%
\bibitem [{\citenamefont {Florijn}\ \emph {et~al.}(2014)\citenamefont
  {Florijn}, \citenamefont {Coulais},\ and\ \citenamefont {van
  Hecke}}]{florijn2014programmable}%
  \BibitemOpen
  \bibfield  {author} {\bibinfo {author} {\bibfnamefont {B.}~\bibnamefont
  {Florijn}}, \bibinfo {author} {\bibfnamefont {C.}~\bibnamefont {Coulais}},\
  and\ \bibinfo {author} {\bibfnamefont {M.}~\bibnamefont {van Hecke}},\
  }\bibfield  {title} {\bibinfo {title} {Programmable mechanical
  metamaterials},\ }\href@noop {} {\bibfield  {journal} {\bibinfo  {journal}
  {Physical review letters}\ }\textbf {\bibinfo {volume} {113}},\ \bibinfo
  {pages} {175503} (\bibinfo {year} {2014})}\BibitemShut {NoStop}%
\bibitem [{\citenamefont {Restrepo}\ \emph {et~al.}(2015)\citenamefont
  {Restrepo}, \citenamefont {Mankame},\ and\ \citenamefont
  {Zavattieri}}]{restrepo2015phase}%
  \BibitemOpen
  \bibfield  {author} {\bibinfo {author} {\bibfnamefont {D.}~\bibnamefont
  {Restrepo}}, \bibinfo {author} {\bibfnamefont {N.~D.}\ \bibnamefont
  {Mankame}},\ and\ \bibinfo {author} {\bibfnamefont {P.~D.}\ \bibnamefont
  {Zavattieri}},\ }\bibfield  {title} {\bibinfo {title} {Phase transforming
  cellular materials},\ }\href@noop {} {\bibfield  {journal} {\bibinfo
  {journal} {Extreme Mechanics Letters}\ }\textbf {\bibinfo {volume} {4}},\
  \bibinfo {pages} {52} (\bibinfo {year} {2015})}\BibitemShut {NoStop}%
\bibitem [{\citenamefont {Shan}\ \emph {et~al.}(2015)\citenamefont {Shan},
  \citenamefont {Kang}, \citenamefont {Raney}, \citenamefont {Wang},
  \citenamefont {Fang}, \citenamefont {Candido}, \citenamefont {Lewis},\ and\
  \citenamefont {Bertoldi}}]{Shan_AdvMat2015}%
  \BibitemOpen
  \bibfield  {author} {\bibinfo {author} {\bibfnamefont {S.}~\bibnamefont
  {Shan}}, \bibinfo {author} {\bibfnamefont {S.~H.}\ \bibnamefont {Kang}},
  \bibinfo {author} {\bibfnamefont {J.~R.}\ \bibnamefont {Raney}}, \bibinfo
  {author} {\bibfnamefont {P.}~\bibnamefont {Wang}}, \bibinfo {author}
  {\bibfnamefont {L.}~\bibnamefont {Fang}}, \bibinfo {author} {\bibfnamefont
  {F.}~\bibnamefont {Candido}}, \bibinfo {author} {\bibfnamefont {J.~A.}\
  \bibnamefont {Lewis}},\ and\ \bibinfo {author} {\bibfnamefont
  {K.}~\bibnamefont {Bertoldi}},\ }\bibfield  {title} {\bibinfo {title}
  {Multistable architected materials for trapping elastic strain energy},\
  }\href {https://doi.org/10.1002/adma.201501708} {\bibfield  {journal}
  {\bibinfo  {journal} {Adv. Mater.}\ }\textbf {\bibinfo {volume} {27}},\
  \bibinfo {pages} {4296} (\bibinfo {year} {2015})}\BibitemShut {NoStop}%
\bibitem [{\citenamefont {Frenzel}\ \emph {et~al.}(2016)\citenamefont
  {Frenzel}, \citenamefont {Findeisen}, \citenamefont {Kadic}, \citenamefont
  {Gumbsch},\ and\ \citenamefont {Wegener}}]{frenzel2016tailored}%
  \BibitemOpen
  \bibfield  {author} {\bibinfo {author} {\bibfnamefont {T.}~\bibnamefont
  {Frenzel}}, \bibinfo {author} {\bibfnamefont {C.}~\bibnamefont {Findeisen}},
  \bibinfo {author} {\bibfnamefont {M.}~\bibnamefont {Kadic}}, \bibinfo
  {author} {\bibfnamefont {P.}~\bibnamefont {Gumbsch}},\ and\ \bibinfo {author}
  {\bibfnamefont {M.}~\bibnamefont {Wegener}},\ }\bibfield  {title} {\bibinfo
  {title} {Tailored buckling microlattices as reusable light-weight shock
  absorbers},\ }\href@noop {} {\bibfield  {journal} {\bibinfo  {journal}
  {Advanced Materials}\ }\textbf {\bibinfo {volume} {28}},\ \bibinfo {pages}
  {5865} (\bibinfo {year} {2016})}\BibitemShut {NoStop}%
\bibitem [{\citenamefont {Yuan}\ \emph {et~al.}(2019)\citenamefont {Yuan},
  \citenamefont {Chua},\ and\ \citenamefont {Zhou}}]{yuan20193d}%
  \BibitemOpen
  \bibfield  {author} {\bibinfo {author} {\bibfnamefont {S.}~\bibnamefont
  {Yuan}}, \bibinfo {author} {\bibfnamefont {C.~K.}\ \bibnamefont {Chua}},\
  and\ \bibinfo {author} {\bibfnamefont {K.}~\bibnamefont {Zhou}},\ }\bibfield
  {title} {\bibinfo {title} {3d-printed mechanical metamaterials with high
  energy absorption},\ }\href@noop {} {\bibfield  {journal} {\bibinfo
  {journal} {Advanced Materials Technologies}\ }\textbf {\bibinfo {volume}
  {4}},\ \bibinfo {pages} {1800419} (\bibinfo {year} {2019})}\BibitemShut
  {NoStop}%
\bibitem [{\citenamefont {Dykstra}\ \emph {et~al.}(2019)\citenamefont
  {Dykstra}, \citenamefont {Busink}, \citenamefont {Ennis},\ and\ \citenamefont
  {Coulais}}]{dykstra2019viscoelastic}%
  \BibitemOpen
  \bibfield  {author} {\bibinfo {author} {\bibfnamefont {D.~M.}\ \bibnamefont
  {Dykstra}}, \bibinfo {author} {\bibfnamefont {J.}~\bibnamefont {Busink}},
  \bibinfo {author} {\bibfnamefont {B.}~\bibnamefont {Ennis}},\ and\ \bibinfo
  {author} {\bibfnamefont {C.}~\bibnamefont {Coulais}},\ }\bibfield  {title}
  {\bibinfo {title} {Viscoelastic snapping metamaterials},\ }\href@noop {}
  {\bibfield  {journal} {\bibinfo  {journal} {Journal of Applied Mechanics}\
  }\textbf {\bibinfo {volume} {86}} (\bibinfo {year} {2019})}\BibitemShut
  {NoStop}%
\bibitem [{\citenamefont {Hector}\ \emph {et~al.}(2023)\citenamefont {Hector},
  \citenamefont {Jarrold}, \citenamefont {Cho}, \citenamefont {Restrepo},
  \citenamefont {Mankame},\ and\ \citenamefont
  {Zavattieri}}]{hector2023energy}%
  \BibitemOpen
  \bibfield  {author} {\bibinfo {author} {\bibfnamefont {K.~W.}\ \bibnamefont
  {Hector}}, \bibinfo {author} {\bibfnamefont {G.}~\bibnamefont {Jarrold}},
  \bibinfo {author} {\bibfnamefont {Y.}~\bibnamefont {Cho}}, \bibinfo {author}
  {\bibfnamefont {D.}~\bibnamefont {Restrepo}}, \bibinfo {author}
  {\bibfnamefont {N.}~\bibnamefont {Mankame}},\ and\ \bibinfo {author}
  {\bibfnamefont {P.~D.}\ \bibnamefont {Zavattieri}},\ }\bibfield  {title}
  {\bibinfo {title} {Energy dissipating architected materials with transversely
  curved tapes and independently tunable properties},\ }\href@noop {}
  {\bibfield  {journal} {\bibinfo  {journal} {Extreme Mechanics Letters}\
  }\textbf {\bibinfo {volume} {58}},\ \bibinfo {pages} {101946} (\bibinfo
  {year} {2023})}\BibitemShut {NoStop}%
\bibitem [{\citenamefont {Grima}\ and\ \citenamefont
  {Evans}(2000)}]{grima_auxetic}%
  \BibitemOpen
  \bibfield  {author} {\bibinfo {author} {\bibfnamefont {J.}~\bibnamefont
  {Grima}}\ and\ \bibinfo {author} {\bibfnamefont {K.}~\bibnamefont {Evans}},\
  }\bibfield  {title} {\bibinfo {title} {Auxetic behavior from rotating
  squares},\ }\href@noop {} {\bibfield  {journal} {\bibinfo  {journal} {J. Mat.
  Sci. Let.}\ }\textbf {\bibinfo {volume} {19}},\ \bibinfo {pages} {1563}
  (\bibinfo {year} {2000})}\BibitemShut {NoStop}%
\bibitem [{\citenamefont {Mullin}\ \emph {et~al.}(2007)\citenamefont {Mullin},
  \citenamefont {Deschanel}, \citenamefont {Bertoldi},\ and\ \citenamefont
  {Boyce}}]{Mullin_PRL2007}%
  \BibitemOpen
  \bibfield  {author} {\bibinfo {author} {\bibfnamefont {T.}~\bibnamefont
  {Mullin}}, \bibinfo {author} {\bibfnamefont {S.}~\bibnamefont {Deschanel}},
  \bibinfo {author} {\bibfnamefont {K.}~\bibnamefont {Bertoldi}},\ and\
  \bibinfo {author} {\bibfnamefont {M.}~\bibnamefont {Boyce}},\ }\bibfield
  {title} {\bibinfo {title} {Pattern transformation triggered by deformation},\
  }\href@noop {} {\bibfield  {journal} {\bibinfo  {journal} {Phys. Rev. Lett.}\
  }\textbf {\bibinfo {volume} {99}} (\bibinfo {year} {2007})}\BibitemShut
  {NoStop}%
\bibitem [{\citenamefont {Bertoldi}\ \emph {et~al.}(2017)\citenamefont
  {Bertoldi}, \citenamefont {Vitelli}, \citenamefont {Christensen},\ and\
  \citenamefont {van Hecke}}]{Bertoldi_NatRevMat}%
  \BibitemOpen
  \bibfield  {author} {\bibinfo {author} {\bibfnamefont {K.}~\bibnamefont
  {Bertoldi}}, \bibinfo {author} {\bibfnamefont {V.}~\bibnamefont {Vitelli}},
  \bibinfo {author} {\bibfnamefont {J.}~\bibnamefont {Christensen}},\ and\
  \bibinfo {author} {\bibfnamefont {M.}~\bibnamefont {van Hecke}},\ }\bibfield
  {title} {\bibinfo {title} {Flexible mechanical metamaterials},\ }\href
  {https://doi.org/10.1038/natrevmats.2017.66} {\bibfield  {journal} {\bibinfo
  {journal} {Nat. Rev. Mater.}\ }\textbf {\bibinfo {volume} {2}},\ \bibinfo
  {pages} {17066} (\bibinfo {year} {2017})}\BibitemShut {NoStop}%
\bibitem [{\citenamefont {Inman}\ and\ \citenamefont
  {Singh}(1994)}]{inman1994engineering}%
  \BibitemOpen
  \bibfield  {author} {\bibinfo {author} {\bibfnamefont {D.~J.}\ \bibnamefont
  {Inman}}\ and\ \bibinfo {author} {\bibfnamefont {R.~C.}\ \bibnamefont
  {Singh}},\ }\href@noop {} {\emph {\bibinfo {title} {Engineering
  vibration}}},\ Vol.~\bibinfo {volume} {3}\ (\bibinfo  {publisher} {Prentice
  Hall Englewood Cliffs, NJ},\ \bibinfo {year} {1994})\BibitemShut {NoStop}%
\bibitem [{\citenamefont {Bertoldi}\ \emph {et~al.}(2010)\citenamefont
  {Bertoldi}, \citenamefont {Reis}, \citenamefont {Willshaw},\ and\
  \citenamefont {Mullin}}]{bertoldi2010negative}%
  \BibitemOpen
  \bibfield  {author} {\bibinfo {author} {\bibfnamefont {K.}~\bibnamefont
  {Bertoldi}}, \bibinfo {author} {\bibfnamefont {P.~M.}\ \bibnamefont {Reis}},
  \bibinfo {author} {\bibfnamefont {S.}~\bibnamefont {Willshaw}},\ and\
  \bibinfo {author} {\bibfnamefont {T.}~\bibnamefont {Mullin}},\ }\bibfield
  {title} {\bibinfo {title} {Negative poisson's ratio behavior induced by an
  elastic instability},\ }\href@noop {} {\bibfield  {journal} {\bibinfo
  {journal} {Advanced materials}\ }\textbf {\bibinfo {volume} {22}},\ \bibinfo
  {pages} {361} (\bibinfo {year} {2010})}\BibitemShut {NoStop}%
\bibitem [{\citenamefont {Kovacic}\ and\ \citenamefont
  {Brennan}(2011)}]{kovacic2011duffing}%
  \BibitemOpen
  \bibfield  {author} {\bibinfo {author} {\bibfnamefont {I.}~\bibnamefont
  {Kovacic}}\ and\ \bibinfo {author} {\bibfnamefont {M.~J.}\ \bibnamefont
  {Brennan}},\ }\href@noop {} {\emph {\bibinfo {title} {The Duffing equation:
  nonlinear oscillators and their behaviour}}}\ (\bibinfo  {publisher} {John
  Wiley \& Sons},\ \bibinfo {year} {2011})\BibitemShut {NoStop}%
\bibitem [{\citenamefont {Brennan}\ \emph {et~al.}(2008)\citenamefont
  {Brennan}, \citenamefont {Kovacic}, \citenamefont {Carrella},\ and\
  \citenamefont {Waters}}]{brennan2008jump}%
  \BibitemOpen
  \bibfield  {author} {\bibinfo {author} {\bibfnamefont {M.}~\bibnamefont
  {Brennan}}, \bibinfo {author} {\bibfnamefont {I.}~\bibnamefont {Kovacic}},
  \bibinfo {author} {\bibfnamefont {A.}~\bibnamefont {Carrella}},\ and\
  \bibinfo {author} {\bibfnamefont {T.}~\bibnamefont {Waters}},\ }\bibfield
  {title} {\bibinfo {title} {On the jump-up and jump-down frequencies of the
  duffing oscillator},\ }\href@noop {} {\bibfield  {journal} {\bibinfo
  {journal} {Journal of Sound and Vibration}\ }\textbf {\bibinfo {volume}
  {318}},\ \bibinfo {pages} {1250} (\bibinfo {year} {2008})}\BibitemShut
  {NoStop}%
\bibitem [{\citenamefont {Guckenheimer}\ and\ \citenamefont
  {Holmes}(2013)}]{guckenheimer2013nonlinear}%
  \BibitemOpen
  \bibfield  {author} {\bibinfo {author} {\bibfnamefont {J.}~\bibnamefont
  {Guckenheimer}}\ and\ \bibinfo {author} {\bibfnamefont {P.}~\bibnamefont
  {Holmes}},\ }\href@noop {} {\emph {\bibinfo {title} {Nonlinear oscillations,
  dynamical systems, and bifurcations of vector fields}}},\ Vol.~\bibinfo
  {volume} {42}\ (\bibinfo  {publisher} {Springer Science \& Business Media},\
  \bibinfo {year} {2013})\BibitemShut {NoStop}%
\bibitem [{\citenamefont {Dykstra}\ \emph {et~al.}(2022)\citenamefont
  {Dykstra}, \citenamefont {Janbaz},\ and\ \citenamefont
  {Coulais}}]{dykstra2022extreme}%
  \BibitemOpen
  \bibfield  {author} {\bibinfo {author} {\bibfnamefont {D.~M.}\ \bibnamefont
  {Dykstra}}, \bibinfo {author} {\bibfnamefont {S.}~\bibnamefont {Janbaz}},\
  and\ \bibinfo {author} {\bibfnamefont {C.}~\bibnamefont {Coulais}},\
  }\bibfield  {title} {\bibinfo {title} {The extreme mechanics of viscoelastic
  metamaterials},\ }\href@noop {} {\bibfield  {journal} {\bibinfo  {journal}
  {APL Materials}\ }\textbf {\bibinfo {volume} {10}},\ \bibinfo {pages}
  {080702} (\bibinfo {year} {2022})}\BibitemShut {NoStop}%
\bibitem [{\citenamefont {Ashby}(2016)}]{Ashby2016}%
  \BibitemOpen
  \bibfield  {author} {\bibinfo {author} {\bibfnamefont {M.}~\bibnamefont
  {Ashby}},\ }\href@noop {} {\emph {\bibinfo {title} {Materials Selection in
  Mechanical Design}}},\ \bibinfo {edition} {5th}\ ed.\ (\bibinfo  {publisher}
  {Elsevier},\ \bibinfo {year} {2016})\BibitemShut {NoStop}%
\bibitem [{\citenamefont {Lakes}(2001)}]{lakes2001extreme_PRL}%
  \BibitemOpen
  \bibfield  {author} {\bibinfo {author} {\bibfnamefont {R.}~\bibnamefont
  {Lakes}},\ }\bibfield  {title} {\bibinfo {title} {Extreme damping in
  composite materials with a negative stiffness phase},\ }\href@noop {}
  {\bibfield  {journal} {\bibinfo  {journal} {Physical review letters}\
  }\textbf {\bibinfo {volume} {86}},\ \bibinfo {pages} {2897} (\bibinfo {year}
  {2001})}\BibitemShut {NoStop}%
\bibitem [{\citenamefont {Masurier}(2021)}]{masurier2021thesis}%
  \BibitemOpen
  \bibfield  {author} {\bibinfo {author} {\bibfnamefont {A.}~\bibnamefont
  {Masurier}},\ }\href@noop {} {\emph {\bibinfo {title} {Dissipative
  Metamaterials for vibration damping}}},\ \bibinfo {type} {Tech. Rep.}\
  (\bibinfo  {institution} {ENSTA Paris},\ \bibinfo {year} {2021})\BibitemShut
  {NoStop}%
\bibitem [{\citenamefont {Rafsanjani}\ and\ \citenamefont
  {Bertoldi}(2017)}]{rafsanjani2017buckling}%
  \BibitemOpen
  \bibfield  {author} {\bibinfo {author} {\bibfnamefont {A.}~\bibnamefont
  {Rafsanjani}}\ and\ \bibinfo {author} {\bibfnamefont {K.}~\bibnamefont
  {Bertoldi}},\ }\bibfield  {title} {\bibinfo {title} {Buckling-induced
  kirigami},\ }\href@noop {} {\bibfield  {journal} {\bibinfo  {journal}
  {Physical review letters}\ }\textbf {\bibinfo {volume} {118}},\ \bibinfo
  {pages} {084301} (\bibinfo {year} {2017})}\BibitemShut {NoStop}%
\bibitem [{\citenamefont {Vella}(2019)}]{vella2019buffering}%
  \BibitemOpen
  \bibfield  {author} {\bibinfo {author} {\bibfnamefont {D.}~\bibnamefont
  {Vella}},\ }\bibfield  {title} {\bibinfo {title} {Buffering by buckling as a
  route for elastic deformation},\ }\href@noop {} {\bibfield  {journal}
  {\bibinfo  {journal} {Nature Reviews Physics}\ }\textbf {\bibinfo {volume}
  {1}},\ \bibinfo {pages} {425} (\bibinfo {year} {2019})}\BibitemShut {NoStop}%
\bibitem [{\citenamefont {Chen}\ \emph {et~al.}(2020)\citenamefont {Chen},
  \citenamefont {Chen}, \citenamefont {Zhang}, \citenamefont {Li},\ and\
  \citenamefont {Li}}]{chen2020kirigami}%
  \BibitemOpen
  \bibfield  {author} {\bibinfo {author} {\bibfnamefont {S.}~\bibnamefont
  {Chen}}, \bibinfo {author} {\bibfnamefont {J.}~\bibnamefont {Chen}}, \bibinfo
  {author} {\bibfnamefont {X.}~\bibnamefont {Zhang}}, \bibinfo {author}
  {\bibfnamefont {Z.-Y.}\ \bibnamefont {Li}},\ and\ \bibinfo {author}
  {\bibfnamefont {J.}~\bibnamefont {Li}},\ }\bibfield  {title} {\bibinfo
  {title} {Kirigami/origami: unfolding the new regime of advanced 3d
  microfabrication/nanofabrication with “folding”},\ }\href@noop {}
  {\bibfield  {journal} {\bibinfo  {journal} {Light: Science \& Applications}\
  }\textbf {\bibinfo {volume} {9}},\ \bibinfo {pages} {1} (\bibinfo {year}
  {2020})}\BibitemShut {NoStop}%
\bibitem [{\citenamefont {Smeets}\ \emph {et~al.}(2021)\citenamefont {Smeets},
  \citenamefont {Fagan}, \citenamefont {Matthews}, \citenamefont {Telford},
  \citenamefont {Murray}, \citenamefont {Pavlov}, \citenamefont {Weafer},
  \citenamefont {Meier},\ and\ \citenamefont {Goggins}}]{smeets2021structural}%
  \BibitemOpen
  \bibfield  {author} {\bibinfo {author} {\bibfnamefont {B.~J.}\ \bibnamefont
  {Smeets}}, \bibinfo {author} {\bibfnamefont {E.~M.}\ \bibnamefont {Fagan}},
  \bibinfo {author} {\bibfnamefont {K.}~\bibnamefont {Matthews}}, \bibinfo
  {author} {\bibfnamefont {R.}~\bibnamefont {Telford}}, \bibinfo {author}
  {\bibfnamefont {B.~R.}\ \bibnamefont {Murray}}, \bibinfo {author}
  {\bibfnamefont {L.}~\bibnamefont {Pavlov}}, \bibinfo {author} {\bibfnamefont
  {B.}~\bibnamefont {Weafer}}, \bibinfo {author} {\bibfnamefont
  {P.}~\bibnamefont {Meier}},\ and\ \bibinfo {author} {\bibfnamefont
  {J.}~\bibnamefont {Goggins}},\ }\bibfield  {title} {\bibinfo {title}
  {Structural testing of a shear web attachment point on a composite lattice
  cylinder for aerospace applications},\ }\href@noop {} {\bibfield  {journal}
  {\bibinfo  {journal} {Composites Part B: Engineering}\ }\textbf {\bibinfo
  {volume} {212}},\ \bibinfo {pages} {108691} (\bibinfo {year}
  {2021})}\BibitemShut {NoStop}%
\bibitem [{\citenamefont {Hunt}\ \emph {et~al.}(2022)\citenamefont {Hunt},
  \citenamefont {Morabito}, \citenamefont {Grace}, \citenamefont {Zhao},\ and\
  \citenamefont {Woods}}]{hunt2022review}%
  \BibitemOpen
  \bibfield  {author} {\bibinfo {author} {\bibfnamefont {C.~J.}\ \bibnamefont
  {Hunt}}, \bibinfo {author} {\bibfnamefont {F.}~\bibnamefont {Morabito}},
  \bibinfo {author} {\bibfnamefont {C.}~\bibnamefont {Grace}}, \bibinfo
  {author} {\bibfnamefont {Y.}~\bibnamefont {Zhao}},\ and\ \bibinfo {author}
  {\bibfnamefont {B.~K.}\ \bibnamefont {Woods}},\ }\bibfield  {title} {\bibinfo
  {title} {A review of composite lattice structures},\ }\href@noop {}
  {\bibfield  {journal} {\bibinfo  {journal} {Composite Structures}\ }\textbf
  {\bibinfo {volume} {284}},\ \bibinfo {pages} {115120} (\bibinfo {year}
  {2022})}\BibitemShut {NoStop}%
\bibitem [{\citenamefont {Deshpande}\ \emph {et~al.}(2001)\citenamefont
  {Deshpande}, \citenamefont {Ashby},\ and\ \citenamefont
  {Fleck}}]{deshpande2001foam}%
  \BibitemOpen
  \bibfield  {author} {\bibinfo {author} {\bibfnamefont {V.}~\bibnamefont
  {Deshpande}}, \bibinfo {author} {\bibfnamefont {M.}~\bibnamefont {Ashby}},\
  and\ \bibinfo {author} {\bibfnamefont {N.}~\bibnamefont {Fleck}},\ }\bibfield
   {title} {\bibinfo {title} {Foam topology: bending versus stretching
  dominated architectures},\ }\href@noop {} {\bibfield  {journal} {\bibinfo
  {journal} {Acta materialia}\ }\textbf {\bibinfo {volume} {49}},\ \bibinfo
  {pages} {1035} (\bibinfo {year} {2001})}\BibitemShut {NoStop}%
\bibitem [{\citenamefont {Diez-Jimenez}\ \emph {et~al.}(2019)\citenamefont
  {Diez-Jimenez}, \citenamefont {Rizzo}, \citenamefont
  {G{\'o}mez-Garc{\'\i}a},\ and\ \citenamefont {Corral-Abad}}]{diez2019review}%
  \BibitemOpen
  \bibfield  {author} {\bibinfo {author} {\bibfnamefont {E.}~\bibnamefont
  {Diez-Jimenez}}, \bibinfo {author} {\bibfnamefont {R.}~\bibnamefont {Rizzo}},
  \bibinfo {author} {\bibfnamefont {M.-J.}\ \bibnamefont
  {G{\'o}mez-Garc{\'\i}a}},\ and\ \bibinfo {author} {\bibfnamefont
  {E.}~\bibnamefont {Corral-Abad}},\ }\bibfield  {title} {\bibinfo {title}
  {Review of passive electromagnetic devices for vibration damping and
  isolation},\ }\href@noop {} {\bibfield  {journal} {\bibinfo  {journal} {Shock
  and Vibration}\ }\textbf {\bibinfo {volume} {2019}} (\bibinfo {year}
  {2019})}\BibitemShut {NoStop}%
\bibitem [{\citenamefont {Yan}\ \emph {et~al.}(2017)\citenamefont {Yan},
  \citenamefont {Wang}, \citenamefont {Hu}, \citenamefont {Wu},\ and\
  \citenamefont {Zhang}}]{yan2017shunt}%
  \BibitemOpen
  \bibfield  {author} {\bibinfo {author} {\bibfnamefont {B.}~\bibnamefont
  {Yan}}, \bibinfo {author} {\bibfnamefont {K.}~\bibnamefont {Wang}}, \bibinfo
  {author} {\bibfnamefont {Z.}~\bibnamefont {Hu}}, \bibinfo {author}
  {\bibfnamefont {C.}~\bibnamefont {Wu}},\ and\ \bibinfo {author}
  {\bibfnamefont {X.}~\bibnamefont {Zhang}},\ }\bibfield  {title} {\bibinfo
  {title} {Shunt damping vibration control technology: A review},\ }\href@noop
  {} {\bibfield  {journal} {\bibinfo  {journal} {Applied Sciences}\ }\textbf
  {\bibinfo {volume} {7}},\ \bibinfo {pages} {494} (\bibinfo {year}
  {2017})}\BibitemShut {NoStop}%
\bibitem [{\citenamefont {Gripp}\ and\ \citenamefont
  {Rade}(2018)}]{gripp2018vibration}%
  \BibitemOpen
  \bibfield  {author} {\bibinfo {author} {\bibfnamefont {J.}~\bibnamefont
  {Gripp}}\ and\ \bibinfo {author} {\bibfnamefont {D.}~\bibnamefont {Rade}},\
  }\bibfield  {title} {\bibinfo {title} {Vibration and noise control using
  shunted piezoelectric transducers: A review},\ }\href@noop {} {\bibfield
  {journal} {\bibinfo  {journal} {Mechanical Systems and Signal Processing}\
  }\textbf {\bibinfo {volume} {112}},\ \bibinfo {pages} {359} (\bibinfo {year}
  {2018})}\BibitemShut {NoStop}%
\bibitem [{\citenamefont {Bossart}\ \emph {et~al.}(2021)\citenamefont
  {Bossart}, \citenamefont {Dykstra}, \citenamefont {van~der Laan},\ and\
  \citenamefont {Coulais}}]{bossart2021oligomodal}%
  \BibitemOpen
  \bibfield  {author} {\bibinfo {author} {\bibfnamefont {A.}~\bibnamefont
  {Bossart}}, \bibinfo {author} {\bibfnamefont {D.~M.}\ \bibnamefont
  {Dykstra}}, \bibinfo {author} {\bibfnamefont {J.}~\bibnamefont {van~der
  Laan}},\ and\ \bibinfo {author} {\bibfnamefont {C.}~\bibnamefont {Coulais}},\
  }\bibfield  {title} {\bibinfo {title} {Oligomodal metamaterials with
  multifunctional mechanics},\ }\href@noop {} {\bibfield  {journal} {\bibinfo
  {journal} {Proceedings of the National Academy of Sciences}\ }\textbf
  {\bibinfo {volume} {118}} (\bibinfo {year} {2021})}\BibitemShut {NoStop}%
\bibitem [{\citenamefont {Shaw}\ and\ \citenamefont
  {Holmes}(1983)}]{shaw1983periodically}%
  \BibitemOpen
  \bibfield  {author} {\bibinfo {author} {\bibfnamefont {S.~W.}\ \bibnamefont
  {Shaw}}\ and\ \bibinfo {author} {\bibfnamefont {P.}~\bibnamefont {Holmes}},\
  }\bibfield  {title} {\bibinfo {title} {A periodically forced piecewise linear
  oscillator},\ }\href@noop {} {\bibfield  {journal} {\bibinfo  {journal}
  {Journal of sound and vibration}\ }\textbf {\bibinfo {volume} {90}},\
  \bibinfo {pages} {129} (\bibinfo {year} {1983})}\BibitemShut {NoStop}%
\bibitem [{\citenamefont {Jung}\ \emph {et~al.}(2014)\citenamefont {Jung},
  \citenamefont {Epureanu} \emph {et~al.}}]{jung2014nonlinear}%
  \BibitemOpen
  \bibfield  {author} {\bibinfo {author} {\bibfnamefont {C.}~\bibnamefont
  {Jung}}, \bibinfo {author} {\bibfnamefont {B.~I.}\ \bibnamefont {Epureanu}},
  \emph {et~al.},\ }\bibfield  {title} {\bibinfo {title} {Nonlinear amplitude
  approximation for bilinear systems},\ }\href@noop {} {\bibfield  {journal}
  {\bibinfo  {journal} {Journal of Sound and Vibration}\ }\textbf {\bibinfo
  {volume} {333}},\ \bibinfo {pages} {2909} (\bibinfo {year}
  {2014})}\BibitemShut {NoStop}%
\bibitem [{\citenamefont {Irvine}(2009)}]{irvine2009introduction}%
  \BibitemOpen
  \bibfield  {author} {\bibinfo {author} {\bibfnamefont {T.}~\bibnamefont
  {Irvine}},\ }\bibfield  {title} {\bibinfo {title} {An introduction to the
  vibration response spectrum},\ }\href@noop {} {\bibfield  {journal} {\bibinfo
   {journal} {Revision D, Vibrationdata}\ } (\bibinfo {year}
  {2009})}\BibitemShut {NoStop}%
\bibitem [{\citenamefont {Bessa}\ \emph {et~al.}(2019)\citenamefont {Bessa},
  \citenamefont {Glowacki},\ and\ \citenamefont {Houlder}}]{bessa2019bayesian}%
  \BibitemOpen
  \bibfield  {author} {\bibinfo {author} {\bibfnamefont {M.~A.}\ \bibnamefont
  {Bessa}}, \bibinfo {author} {\bibfnamefont {P.}~\bibnamefont {Glowacki}},\
  and\ \bibinfo {author} {\bibfnamefont {M.}~\bibnamefont {Houlder}},\
  }\bibfield  {title} {\bibinfo {title} {Bayesian machine learning in
  metamaterial design: Fragile becomes supercompressible},\ }\href@noop {}
  {\bibfield  {journal} {\bibinfo  {journal} {Advanced Materials}\ }\textbf
  {\bibinfo {volume} {31}},\ \bibinfo {pages} {1904845} (\bibinfo {year}
  {2019})}\BibitemShut {NoStop}%
\bibitem [{\citenamefont {Sobol}(1967)}]{sobol1967distribution}%
  \BibitemOpen
  \bibfield  {author} {\bibinfo {author} {\bibfnamefont {I.~M.}\ \bibnamefont
  {Sobol}},\ }\bibfield  {title} {\bibinfo {title} {On the distribution of
  points in a cube and the approximate evaluation of integrals},\ }\href@noop
  {} {\bibfield  {journal} {\bibinfo  {journal} {Zhurnal Vychislitel'noi
  Matematiki i Matematicheskoi Fiziki}\ }\textbf {\bibinfo {volume} {7}},\
  \bibinfo {pages} {784} (\bibinfo {year} {1967})}\BibitemShut {NoStop}%
\bibitem [{\citenamefont {Dykstra}\ \emph {et~al.}(2023)\citenamefont
  {Dykstra}, \citenamefont {Lenting}, \citenamefont {Masurier},\ and\
  \citenamefont {Coulais}}]{Zenodo2023}%
  \BibitemOpen
  \bibfield  {author} {\bibinfo {author} {\bibfnamefont {D.~M.~J.}\
  \bibnamefont {Dykstra}}, \bibinfo {author} {\bibfnamefont {C.}~\bibnamefont
  {Lenting}}, \bibinfo {author} {\bibfnamefont {A.}~\bibnamefont {Masurier}},\
  and\ \bibinfo {author} {\bibfnamefont {C.}~\bibnamefont {Coulais}},\
  }\bibfield  {title} {\bibinfo {title} {Buckling metamaterials for extreme
  vibration damping},\ }\href@noop {} {\bibfield  {journal} {\bibinfo
  {journal} {Zenodo, 10.5281/zenodo.7661958}\ } (\bibinfo {year}
  {2023})}\BibitemShut {NoStop}%
\end{thebibliography}%

\end{document}